\documentclass[epj]{svjour}
\usepackage{graphics}
\usepackage{amssymb}
\begin{document}
\title{Estimates for parameters and characteristics of the confining 
SU(3)-gluonic field in neutral kaons and chiral limit for pseudoscalar 
nonet}
%\subtitle{Do you have a subtitle?\\ If so, write it here}
\author{Yu. P. Goncharov%\inst{1} \and N. E. Firsova\inst{2}% etc
% \thanks is optional - remove next line if not needed
%\thanks{\emph{Present address:} Insert the address here if needed}%
}                     % Do not remove
\offprints{Yu.P. Goncharov}          % Insert a name or remove this line
\institute{Theoretical Group, Experimental Physics Department, 
State Polytechnical University, Sankt-Petersburg 195251, Russia} 
%\and Institute of the Mechanical Engineering Problems, Russian 
%Academy of Sciences, Sankt-Petersburg 199178, Russia}
%
\date{Received: date / Revised version: date}
% The correct dates will be entered by Springer
%
\abstract{
First part of the paper is devoted to applying the confinement mechanism 
proposed earlier by the author to estimate 
the possible parameters of the confining SU(3)-gluonic field in 
neutral kaons. The estimates obtained are consistent with the widths of the electromagnetic 
decays $K^0,\bar{K}^0\to2\gamma$ too. The corresponding estimates of the gluon 
concentrations, electric and magnetic colour field strengths are also adduced 
for the mentioned field at the scales of the mesons under consideration. The 
second part of the paper takes into account the results obtained previously by 
the author to estimate the purely gluonic contribution 
to the masses of all the mesons of pseudoscalar nonet and also to consider
a possible relation with a phenomenological string-like picture of 
confinement. Finally, the problem of masses in particle physics is 
shortly discussed within the framework of approach to
the chiral symmetry breaking in quantum chromodynamics (QCD) 
proposed recently by the author. 
 % end of PACS codes
} %end of abstract
\authorrunning{Yu. P. Goncharov}% \& N. E. Firsova}%
\titlerunning{Chiral limit for pseudoscalar nonet}
\maketitle
\section{Introduction}
The present paper to some degree summarizes our previous ones on studying the 
pseudoscalar meson nonet within the framework of the (quark) confinement 
mechanism proposed earlier by the author. Global strategy may consist in reconsidering the whole 
spectroscopy of both mesons and baryons from the positions of the mentioned 
mechanism so exploring the pseudoscalar meson nonet is in essence only the 
first step in the given direction. But such a study involves concretizing the 
confinement mechanism itself in the generally accepted physical terms and when 
applying the mechanism to concrete hadrons there always arise new physical 
possibilities of interpreting the results obtained which further enriches the 
mechanism from physical point of view. In other words, the proposed 
confinement mechanism should be continuously modified and improved from physical 
positions and the best way of doing so is to study concrete hadrons with its 
help. Let us now shortly outline the main features of the approach suggested. 

In \cite{{Gon01},{Gon051},{Gon052}} for the Dirac-Yang-Mills 
system derived from 
QCD-Lagrangian  an unique family of compatible 
nonperturbative solutions was found and explored, which could pretend to 
decsribing confinement of two quarks. 
The applications of the family to the description of both the heavy quarkonia 
spectra \cite{{Gon03},{Gon08a}} and a number of properties of pions, kaons, 
$\eta$- and $\eta^\prime$-mesons 
\cite{{Gon06},{Gon07a},{Gon07b},{Gon08},{Gon08b}} 
showed that the confinement 
mechanism is qualitatively the same for both light mesons and heavy quarkonia.
At this moment it can be decribed in the following way.

The next main physical reasons underlie linear confinement in the 
mechanism under discussion. The first one is that gluon exchange between 
quarks is realized with the propagator different from the photon-like one, and 
existence and form of such a propagator is a {\em direct} consequence of the 
unique confining 
nonperturbative solutions of the Yang-Mills equations 
\cite{{Gon051},{Gon052}}. The second reason is that, 
owing to the structure of the mentioned propagator, quarks mainly emit and 
interchange the soft gluons so the gluon condensate (a classical gluon field) 
between quarks basically consists of soft gluons (for more details 
see Refs. \cite{{Gon051},{Gon052}}) but, because of the fact that any gluon 
also emits gluons (still softer), the corresponding gluon concentrations 
rapidly become huge and form a linear confining magnetic colour field of 
enormous strengths, which leads to confinement of quarks. This is by virtue of 
the fact that just the magnetic part of the mentioned propagator is responsible 
for a larger portion of gluon concentrations at large distances since the 
magnetic part has stronger infrared singularities than the electric one. 
In the circumstances 
physically the nonlinearity of the Yang-Mills equations effectively vanishes so the 
latter possess the unique nonperturbative confining solutions of the 
Abelian-like form (with the values in Cartan subalgebra of SU(3)-Lie algebra) 
\cite{{Gon051},{Gon052}} which describe 
the gluon condensate under consideration. Moreover, since the overwhelming 
majority of gluons is soft they cannot leave the hadron (meson) until some 
gluons obtain additional energy (due to an external reason) to rush out. So 
we also deal with the confinement of gluons.  

The approach under discussion equips us with the explicit wave functions 
for every two quarks (meson or quarkonium). The wave functions are parametrized 
by a set of 
real constants $a_j, b_j, B_j$ describing the mentioned 
{\em nonperturbative} confining SU(3)-gluonic field (the gluon condensate) and 
they are {\em nonperturbative} modulo square integrable 
solutions of the Dirac equation in the above confining SU(3)-field and also  
depend on $\mu_0$, the reduced
mass of the current masses of quarks forming meson. It is clear that under the 
given approach just constants $a_j, b_j, B_j,\mu_0$ determine all properties 
of any meson (quarkonium), i. e.,  the approach directly appeals to quark 
and gluonic degrees of freedom as should be according to the first principles 
of QCD. Also it is clear that the constants mentioned should be extracted from 
experimental data. 

Such a program has been to a certain extent advanced in 
Refs. \cite{{Gon03},{Gon06},{Gon07a},{Gon07b},{Gon08},{Gon08a},{Gon08b}}. 
Under the circumstances one aim of the present paper is to complete obtaining 
estimates for $a_j, b_j, B_j$ for the nonet of pseudoscalar mesons and we shall 
here consider neutral kaons 
$K^0,\bar{K}^0$. Another aim is to a certain degree to analyse some physical 
conclusions that can be obtained by considering the whole nonet from positions 
of the approach suggested. 

Of course, when conducting our considerations 
we shall rely on the standard quark model (SQM) based on SU(3)-flavor symmetry 
(see, e. g., \cite{pdg}) so in accordance with SQM 
$K^0,\bar{K}^0=d\bar{s}$, $\bar{d}s$ respectively. 

Section 2 contains a survey of main relations underlying 
description of any mesons (quarkonia) in our approach. Section 3 gives 
estimates for parameters of the confining 
SU(3)-gluonic field for neutral kaons and also contains a discussion about 
whether the obtained estimates might also be consistent with the widths of 
2-photon decays $K^0$, $\bar{K}^0$ $\to2\gamma$. Section 4 employs the obtained 
parameters of SU(3)-gluonic field to get the corresponding estimates for such 
characteristics of the mentioned field as gluon concentrations, electric and 
magnetic colour field strengths at the scales of the mesons in question 
while Section 5 deals with discussion about chiral limit for the nonet of 
pseudoscalar mesons. Section 6 takes into account the results obtained 
previously by the author to consider a possible relation with a 
phenomenological string-like picture of confinement. In section 7 
the problem of masses in particle physics is shortly discussed within the 
framework of approach to the chiral symmetry breaking in QCD proposed recently 
by the author. Section 8 is devoted to the concluding remarks. 

Appendices A and B contain the detailed description 
of main building blocks for meson wave functions in the approach under 
discussion, respectively: eigenspinors of the Euclidean Dirac operator on 
two-sphere ${\Bbb S}^2$ and radial parts for the modulo square integrable 
solutions of Dirac equation in the confining SU(3)-Yang-Mills field. At last, 
Appendix C supplements Section 2 with a proof of the uniqueness theorem from 
that Section in the case of SU(3)-Yang-Mills equations. 
 
Further we shall deal with the metric of
the flat Minkowski spacetime $M$ that
we write down (using the ordinary set of local spherical coordinates
$r,\vartheta,\varphi$ for the spatial part) in the form
$$ds^2=g_{\mu\nu}dx^\mu\otimes dx^\nu\equiv
dt^2-dr^2-r^2(d\vartheta^2+\sin^2\vartheta d\varphi^2)\>, \eqno(1)$$
so we have $|\delta|=|\det(g_{\mu\nu})|=(r^2\sin\vartheta)^2$
and $0\leq r<\infty$, $0\leq\vartheta<\pi$,
$0\leq\varphi<2\pi$.

Throughout the paper we employ the Heaviside-Lorentz system of units 
with $\hbar=c=1$, unless explicitly stated otherwise, so the gauge coupling 
constant $g$ and the strong coupling constant ${\alpha_s}$ are connected by 
the relation $g^2/(4\pi)=\alpha_s$. 

When calculating we apply the 
relations $1\ {\rm GeV^{-1}}\approx0.1973269679\ {\rm fm}\>$,
$1\ {\rm s^{-1}}\approx0.658211915\times10^{-24}\ {\rm GeV}\>$, 
$1\ {\rm V/m}\approx0.2309956375\times 10^{-23}\ {\rm GeV}^2$, 
$1\ {\rm T}=4\pi\times10^{-7} {\rm H/m}\times1\ {\rm A/m}
\approx0.6925075988\times 10^{-15}\ {\rm GeV}^2 $. 

Finally, for the necessary estimates we shall employ the $T_{00}$-component 
(volumetric energy density ) of the energy-momentum tensor for a 
SU(3)-Yang-Mills field which should be written in the chosen system of units 
in the form
$$T_{\mu\nu}=-F^a_{\mu\alpha}\,F^a_{\nu\beta}\,g^{\alpha\beta}+
{1\over4}F^a_{\beta\gamma}\,F^a_{\alpha\delta}g^{\alpha\beta}g^{\gamma\delta}
g_{\mu\nu}\>. \eqno(2) $$
\section{Survey of main relations}
\subsection{The confining SU(3)-gluonic field and meson wave functions}
As was mentioned above, our considerations shall be based on the unique family 
of compatible nonperturbative solutions for 
the Dirac-Yang-Mills system (derived from QCD-Lagrangian) studied at the whole 
length in Refs. \cite{{Gon01},{Gon051},{Gon052}}.  Referring for more details 
to those references, let us briefly decribe and specify only the relations 
necessary to us in the present paper. 

One part of the mentioned family is presented by the unique nonperturbative 
confining solution of the SU(3)-Yang-Mills 
equations for the gluonic field $A=A_\mu dx^\mu=
A^a_\mu \lambda_adx^\mu$ ($\lambda_a$ are the 
known Gell-Mann matrices, $\mu=t,r,\vartheta,\varphi$, $a=1,...,8$) and looks 
as follows 
$$ {\mathcal A}_{1t}\equiv A^3_t+\frac{1}{\sqrt{3}}A^8_t =-\frac{a_1}{r}+A_1 \>,
{\mathcal A}_{2t}\equiv -A^3_t+\frac{1}{\sqrt{3}}A^8_t=-\frac{a_2}{r}+A_2\>,$$
$${\mathcal A}_{3t}\equiv-\frac{2}{\sqrt{3}}A^8_t=\frac{a_1+a_2}{r}-(A_1+A_2)\>, $$
$$ {\mathcal A}_{1\varphi}\equiv A^3_\varphi+\frac{1}{\sqrt{3}}A^8_\varphi=
b_1r+B_1 \>,
{\mathcal A}_{2\varphi}\equiv -A^3_\varphi+\frac{1}{\sqrt{3}}A^8_\varphi=
b_2r+B_2\>,$$
$${\mathcal A}_{3\varphi}\equiv-\frac{2}{\sqrt{3}}A^8_\varphi=
-(b_1+b_2)r-(B_1+B_2)\> \eqno(3)$$
with the real constants $a_j, A_j, b_j, B_j$ parametrizing the family. 

The word {\em unique} should be understood in the strict mathematical sense. 
In fact in Ref. \cite{Gon051} the following theorem was proved (see also 
Appendix C):

{\em The unique exact spherically symmetric (nonperturbative) confining 
solutions (depending only on $r$ and $r^{-1}$) of SU(3)-Yang-Mills 
equations in Minkowski spacetime consist of the family of (3)}.

It should be noted that solution (3) was found early in 
Ref. \cite{Gon01} but its uniqueness was proved just in Ref. \cite{Gon051} 
(see also Ref. \cite{Gon052}). Besides, in Ref. \cite{Gon051} (see also 
Ref. \cite{Gon06}) it was shown that the above unique confining solutions (3) 
satisfy the so-called Wilson confinement criterion \cite{Wil}. Up to now 
nobody contested the above results so if we want to describe interaction 
between quarks by spherically symmetric SU(3)-fields then they can be only 
those from the above theorem. On the other hand, the desirability of 
spherically symmetric (colour) interaction between quarks at all distances 
naturally follows from analysing the $p\bar{p}$-collisions (see, e.g., 
Ref. \cite{Per}) where one observes a Coulomb-like potential in events which 
can be identified with scattering quarks on each other, i.e., actually at small 
distances one observes the Coulomb-like part of solution (3). Under 
this situation, a natural assumption will be that the quark interaction remains 
spherically symmetric at large distances too but then, if trying to extend 
the Coulomb-like part to large distances in a spherically symmetric way, we 
shall inevitably come to the solution (3) in virtue of the above theorem.  

Now one should say that the similar unique confining solutions exist for all 
semisimple and non-semisimple compact Lie groups, in particular, for SU($N$) 
with $N\ge2$ and 
U($N$) with $N\ge1$ \cite{{Gon051},{Gon052}}. Explicit form of solutions, 
e.g., for SU($N$) with $N=2,4$ can be found in Ref.\cite{Gon052} but it 
should be emphasized that components linear in $r$ always represent the 
magnetic (colour) field in all the mentioned solutions. Especially, the case 
of the U(1)-group is interesting which corresponds to usual electrodynamics. 
Under this situation, as was pointed out in 
Refs. \cite{{Gon051},{Gon052}}, there is an interesting possibility of 
indirect experimental verification of the confinement mechanism under 
discussion. Indeed the confining solutions 
of Maxwell equations for classical electrodynamics point out 
the confinement phase could be in electrodynamics as well. Though 
there exist no elementary charged particles generating a constant magnetic 
field linear in $r$, the distance from particle, after all, if it could 
generate this elecromagnetic field configuration in laboratory then one might 
study motion of the charged particles in that field. The confining properties 
of the mentioned field should be displayed at classical level too but the exact 
behaviour of particles in this field requires certain analysis of the corresponding 
classical equations of motion. Such a program has been recently realized in 
Ref. \cite{GF10}. Motion of a charged (classical) particle was studied in the 
field representing magnetic part of the mentioned solution of Maxwell equations 
and it was shown that one deals with the full classical confinement of the 
charged particle in such a field: under any initial conditions the particle 
motion is accomplished within a finite region of space so that the particle 
trajectory is near magnetic field lines while the latter are compact manifolds 
(circles). Those results might be useful in thermonuclear plasma physics 
(for more details see \cite{GF10}). 

As has been repeatedly explained in 
Refs. \cite{{Gon051},{Gon052},{Gon03},{Gon06}}, parameters $A_{1,2}$ of 
solution (3) are inessential for physics in question and we can 
consider $A_1=A_2=0$. 
Also, as has been repeatedly discussed by us earlier (see, e. g., 
Refs. \cite{{Gon051},{Gon052}}), from the above form it is clear that 
the solution (3) is a configuration describing the electric Coulomb-like colour 
field (components $A^{3,8}_t$) and the magnetic colour field linear in $r$ 
(components $A^{3,8}_\varphi$) and we wrote down
the solution (3) in the combinations that are just 
needed to insert into the corresponding Dirac equation. 

Another part of the family represents the meson wave functions and is given by 
the unique nonperturbative modulo 
square integrable solutions of the mentioned Dirac equation in the confining 
SU(3)-field of (3) $\Psi=(\Psi_1, \Psi_2, \Psi_3)$ 
with the four-dimensional Dirac spinors 
$\Psi_j$ representing the $j$th colour component of the meson, 
so $\Psi$ may describe the relative motion (relativistic bound states) of two 
quarks in mesons and is at $j=1,2,3$ (with Pauli matrix $\sigma_1$)  
$$\Psi_j=e^{-i\omega_j t}\psi_j\equiv 
e^{-i\omega_j t}r^{-1}\pmatrix{F_{j1}(r)\Phi_j(\vartheta,\varphi)\cr\
F_{j2}(r)\sigma_1\Phi_j(\vartheta,\varphi)}\>,\eqno(4)$$
with the 2D eigenspinor $\Phi_j=\pmatrix{\Phi_{j1}\cr\Phi_{j2}}$ of the
Euclidean Dirac operator ${\mathcal D}_0$ on the unit sphere ${\mathbb S}^2$, while 
the coordinate $r$ stands for the distance between quarks. 

In this situation, if a meson is composed of quarks $q_{1,2}$ with different flavours then 
the energy spectrum of the meson will be given 
by $\epsilon=m_{q_1}+m_{q_2}+\omega$ with the current quark masses $m_{q_k}$ (
rest energies) of the corresponding quarks and an interaction energy $\omega$. 
On the other hand at $j=1,2,3$
$$\omega_j=\omega_j(n_j,l_j,\lambda_j)=$$ 
$$\scriptsize{\frac{\Lambda_j g^2a_jb_j\pm(n_j+\alpha_j)
\sqrt{(n_j^2+2n_j\alpha_j+\Lambda_j^2)\mu_0^2+g^2b_j^2(n_j^2+2n_j\alpha_j)}}
{n_j^2+2n_j\alpha_j+\Lambda_j^2}}\>\eqno(5)$$
with the gauge coupling constant $g$ while $\mu_0$ is a mass parameter and one 
should consider it to be the reduced mass which is equal to 
$ m_{q_1}m_{q_2}/(m_{q_1}+m_{q_2})$ with the current quark masses $m_{q_k}$ (
rest energies) of the corresponding quarks forming a meson (quarkonium), 
$a_3=-(a_1+a_2)$, $b_3=-(b_1+b_2)$, $B_3=-(B_1+B_2)$, 
$\Lambda_j=\lambda_j-gB_j$, $\alpha_j=\sqrt{\Lambda_j^2-g^2a_j^2}$, 
$n_j=0,1,2,...$, while $\lambda_j=\pm(l_j+1)$ are
the eigenvalues of Euclidean Dirac operator ${\mathcal D}_0$ 
on a unit sphere with $l_j=0,1,2,...$. 

In line with the above we should have $\omega=\omega_1=\omega_2=\omega_3$ in 
energy spectrum $\epsilon=m_{q_1}+m_{q_2}+\omega$ for any meson (quarkonium) 
and this at once imposes two conditions on parameters $a_j,b_j,B_j$ when 
choosing some experimental value for $\epsilon$ at the given current quark 
masses $m_{q_1},m_{q_2}$. 

The general form of the radial parts of (4) can be found, e.g., in 
Appendix B and within the given paper we need only the 
radial parts of (4) at $n_j=0$ 
(the ground state) that are 
$$F_{j1}=C_jP_jr^{\alpha_j}e^{-\beta_jr}\left(1-
\frac{gb_j}{\beta_j}\right), P_j=gb_j+\beta_j, $$
$$F_{j2}=iC_jQ_jr^{\alpha_j}e^{-\beta_jr}\left(1+
\frac{gb_j}{\beta_j}\right), Q_j=\mu_0-\omega_j\eqno(6)$$
with $\beta_j=\sqrt{\mu_0^2-\omega_j^2+g^2b_j^2}$, while $C_j$ is determined 
from the normalization condition
$\int_0^\infty(|F_{j1}|^2+|F_{j2}|^2)dr=\frac{1}{3}$. The 
corresponding eigenspinors of (4) with $\lambda =\pm1$ ($l=0$) are 
$$\lambda=-1: \Phi=\frac{C}{2}\pmatrix{e^{i\frac{\vartheta}{2}}
\cr e^{-i\frac{\vartheta}{2}}\cr}e^{i\varphi/2},\> {\rm or}\>\>
\Phi=\frac{C}{2}\pmatrix{e^{i\frac{\vartheta}{2}}\cr
-e^{-i\frac{\vartheta}{2}}\cr}e^{-i\varphi/2},$$
$$\lambda=1: \Phi=\frac{C}{2}\pmatrix{e^{-i\frac{\vartheta}{2}}\cr
e^{i\frac{\vartheta}{2}}\cr}e^{i\varphi/2}, \> {\rm or}\>\>
\Phi=\frac{C}{2}\pmatrix{-e^{-i\frac{\vartheta}{2}}\cr
e^{i\frac{\vartheta}{2}}\cr}e^{-i\varphi/2} 
\eqno(7) $$
with the coefficient $C=1/\sqrt{2\pi}$ (for more details, see 
Appendix A). 

\subsection{Singularities of solutions}
As is seen from (3), the solutions in question have singularities: electric 
part contains the Coulomb-like singularities while we can rewrite the magnetic 
part in terms of differential 1-forms as $(b_jr+B_j)d\varphi$, $j=$1, 2, 3, 
and then pass on to Cartesian coordinates employing the relations 
$$\varphi=\arctan(y/x),\>
d\varphi=\frac{\partial\varphi}{\partial x}dx+
\frac{\partial\varphi}{\partial y}dy\>$$
which entails
$$(b_jr+B_j)d\varphi=
-\frac{(b_jr+B_j)y}{x^2+y^2}dx+\frac{(b_jr+B_j)x}{x^2+y^2}dy\,,$$
wherefrom it is obvious that the colour magnetic field of (3) 
has the singularities on the $z$-axis. 

It should be noted that an analysis of singularites of YM-potentials requires 
both mathematical and physical considerations and, in general, is different 
from classical and quantum-mechanical point of view. 
In our recent paper \cite{GF10} we gave some analysis of those 
singularities from classical point of view.

But let us now note the following. In classical and quantum electrodynamics (QED) 
it is well known (see e. g. \cite{LL}) that the notion 
of classical electromagnetic field (a photon condensate) generated by a 
charged particle is applicable only at distances much greater than the Compton 
wavelength $\lambda_c=1/m$ for the given particle with mass $m$. Within the QCD 
framework the parameter $\Lambda_{QCD}$ plays a similar part (see, 
e.g., \cite{pdg}). Namely, the notion of classical SU(3)-gluonic field 
( a gluon condensate) is 
not applicable at the distances much less than $1/\Lambda_{QCD}$. 

In this situation, the known singularity of the Coulomb potential in QED
$\Phi=\alpha/r$ at $r=0$ makes the purely mathematical sense since from the 
point of view of QED the photon condensate [huge number of (vitrual) photons] 
described by $\Phi$ exists only at $r>>\lambda_c$ while at $r<\lambda_c$ one 
may only speak about single photons rather than about condensate, i.e., 
the field in classical sense. The same holds true, e.g., for magnetic field 
of a uniformly moving charge where its strength 
$H\sim {\bf v}\times {\bf r}/r^3$, ${\bf v}$ is charge velocity. 

The colour magnetic field (3) under consideration has also the 
singularities on the $z$-axis so its formal mathematical definition domain is 
the manifold ${\mathbb{R}}^3\backslash \{z\}$ with the $z$-axis discarded 
rather than the manifold ${\Bbb{R}}^3$. But we should not forget that we do 
not need only the appropriate solutions of YM-equations to describe 
confinement. The YM-equations are only a part of the Dirac-YM system derived from 
QCD-lagrangian by standard prescription. The wave functions of hadrons (at any 
rate, mesons) are given by the modulo square integrable solutions of the Dirac 
equation (which is the second important part of the above Dirac-YM system) in 
the field (3). But the $j$th colour component of wave function of two quarks 
(meson) (see (6)) in such a field behaves 
as $\psi_j\sim r^{\alpha_j}e^{-g|b_j|r}$, ($\alpha_j>0$), at $|b_j|\to\infty$ 
with $b_j$ characterising the linear colour magnetic field of 
solution (3), $r$ is distance between 
quarks, $j=$1, 2, 3, $b_3=-(b_1+b_2)$. I.e., typical size of hadron is 
$r\sim 1/(g|b_j|)\to0$ and we deal just with confinement and besides we can 
see that those wave functions have {\em no singularities} along the $z$-axis 
and are well defined there, i.e. the wave functions are well defined already on 
the whole ${\Bbb{R}}^3$. But just the wave functions are needed to calculate 
miscellaneous characteristics of mesons (masses, radii and so on). So it is 
clear that at such computations the singularities of YM-potentials in 
questions have no influence on physical results, i.e., those singularities are 
not {\em physical} ones. 

Physically this may mean that at large distances 
nonlinearity of the Yang-Mills equations effectively vanishes so the 
latter possess the unique spherically symmetric nonperturbative confining 
solutions (3) (formally defined on ${\mathbb{R}}^3\backslash \{z\}$) of 
the Abelian-like form (with the values in Cartan subalgebra 
of SU(3)-Lie algebra) which describe the gluon condensate ( a classical gluon 
field) leading to the confinement. 

The situation is practically the same as for the hydrogen atom or positronium: 
the wave functions of those systems (see any textbook on quantum mechanics) are 
well defined at $r=0$ so the known singularity in the Coulomb potential 
at $r=0$ is also unphysical one, as said above. So modelling those 
singularities by some $\delta$-functions (which is possible, as can easy  
show) makes no physical sense from the quantum-mechanical point of view and 
it can give just a suitable method for exploring some {\em classical} problems 
which is done, e.g., in many courses of classical electrodynamics while the 
notion of {\em classical chromodynamics} in fact makes no sense: we can never 
generate the {\em classical} SU(3)-Yang-Mills field and 
{\em classical coloured charged} particles at macroscopic scales.  

On the other hand, just quantum considerations can lead to one more point of 
view on the problem of singularity along $z$-axis of magnetic part for 
solution (3) and it is presented in Section 6 of the paper. 

To summarize, solutions (3) are the unique spherically-symmetric solutions 
of YM-equations, though mathematically being formally defined on 
${\mathbb{R}}^3\backslash \{z\}$, but with the {\em unphysical} singularities 
on the $z$-axis which are inessential from quantum-mechanical point of view.  

\subsection{Nonrelativistic and the weak coupling limits}
It is useful to specify the nonrelativistic limit (when 
$c\to\infty$) for the spectrum (5). For this one should replace 
$g\to g/\sqrt{\hbar c}$, 
$a_j\to a_j/\sqrt{\hbar c}$, $b_j\to b_j\sqrt{\hbar c}$, 
$B_j\to B_j/\sqrt{\hbar c}$ and, expanding (5) in $z=1/c$, we shall get
$$\omega_j(n_j,l_j,\lambda_j)=$$
$$\pm\mu_0c^2\left[1\mp
\frac{g^2a_j^2}{2\hbar^2(n_j+|\lambda_j|)^2}z^2\right]+$$
$$\left[\frac{\lambda_j g^2a_jb_j}{\hbar(n_j+|\lambda_j|)^2}\,
\mp\mu_0\frac{g^3B_ja_j^2f(n_j,\lambda_j)}{\hbar^3(n_j+|\lambda_j|)^{7}}\right]
z\,+O(z^2)\>,\eqno(8)$$
where 
$f(n_j,\lambda_j)=4\lambda_jn_j(n_j^2+\lambda_j^2)+
\frac{|\lambda_j|}{\lambda_j}\left(n_j^{4}+6n_j^2\lambda_j^2+\lambda_j^4
\right)$. 

As is seen from (8), at $c\to\infty$ the contribution of linear magnetic 
colour field (parameters $b_j, B_j$) to the spectrum really vanishes and the 

spectrum in essence becomes the purely nonrelativistic Coulomb one (modulo 
the rest energy). Also it is 
clear that when $n_j\to\infty$, $\omega_j\to\pm\sqrt{\mu_0^2+g^2b_j^2}$. 
At last, one should specify the weak 
coupling limit of (5), i.e., the case $g\to0$. As is not complicated to see 
from (5), $\omega_j\to\pm\mu_0$ when $g\to0$. But then quantities 
$\beta_j=\sqrt{\mu_0^2-\omega_j^2+g^2b_j^2}\to0$ and wave functions of (6) 
cease to be the modulo square integrable ones at $g=0$, i.e., they cease to 
describe relativistic bound states. Accordingly, this means that the equation 
(5) does not make physical meaning at $g=0$. 

We may seemingly use (5) with various combinations of signes ($\pm$) before 
the second summand in numerators of (5) but, due to (8), it is 
reasonable to take all signs equal to plus which is our choice within the 
paper. Besides, 
as is not complicated to see, radial parts in the nonrelativistic limit have 
the behaviour of form $F_{j1},F_{j2}\sim r^{l_j+1}$, which allows one to call 
quantum number $l_j$ angular momentum for the $j$th colour component though 
angular momentum is not conserved in the field (3) \cite{{Gon01},{Gon052}}. So, 
for mesons under consideration we should put all $l_j=0$. 
\subsection{Chiral limit}
There is one more interesting limit for relation (5) -- the chiral one, i.e., 
the situation when $m_{q1},m_{q2}\to0$ which entails $\mu_0\to0$ and (5) reduces 
to (at $j=1,2,3$)
$$(\omega_j)_{\rm chiral}=\frac{\Lambda_j g^2a_jb_j\pm(n_j+\alpha_j)g|b_j|
\sqrt{n_j^2+2n_j\alpha_j}}
{n_j^2+2n_j\alpha_j+\Lambda_j^2}\>, \eqno(9)$$
which mathematically signifies that the Dirac equation in the field (3) 
possesses a nontrivial spectrum of bound states even for massless fermions. 
Physically this gives us  
a possible approach to the problem of chiral symmetry breaking in QCD 
\cite{Gon08b}: in chirally symmetric world masses of mesons are fully 
determined by the confining SU(3)-gluonic field between (massless) quarks 
and not equal to zero. Accordingly chiral symmetry is a sufficiently rough 
approximation holding true only when neglecting the mentioned SU(3)-gluonic 
field between quarks and no additional mechanism of the spontaneous chiral 
symmetry breaking connected to the so-called Goldstone bosons is required. 
As a result, e.g., masses of mesons from pseudoscalar nonet have a purely 
gluonic contribution and we shall consider it in section 5. 

One can note that for being the nonzero chiral limit of (5) the crucial role 
belongs to the colour magnetic field linear in $r$ [parameters $b_{1,2}$ from 
solution (3)] inasmuch as chiral limit is equal exactly to zero when 
$b_{1,2}=0$. On the contrary, when parameters $a_{1,2}$ of the Coulomb colour 
electric part of solution (3) are equal to zero, the chiral limit may be nonzero 
at $b_{1,2}\ne0$, as is seen from (9) except for the case $n_j=0$ when 
both parts of SU(3)-gluonic field (3) are important for confinement and mass 
generation in chiral limit.  

\subsection{Choice of quark masses and the gauge coupling constant}
Obviously, we should choose a few quantities that are the most important from 
the physical point of view to characterize 
mesons under consideration and then we should evaluate the given quantities 
within the framework of our approach. In the circumstances let us settle on 
the ground state energy (mass) of neutral kaons, the root-mean-square 
radius of them and the magnetic moment. All three magnitudes are essentially 
nonperturbative ones, and can be calculated only by nonperturbative techniques.

Within the present paper we shall use relations (5) at $n_j=0=l_j$ so 
energy (mass) of mesons under consideration is given by $\mu=m_d+m_s+\omega$ 
with $\omega=\omega_j(0,0,\lambda_j)$ for any $j=1,2,3$ whereas 
$$\omega=\frac{g^2a_1b_1}{\Lambda_1}+\frac{\alpha_1\mu_0}
{|\Lambda_1|}=\frac{g^2a_2b_2}{\Lambda_2}+\frac{\alpha_2\mu_0}
{|\Lambda_2|}=$$
$$\frac{g^2a_3b_3}{\Lambda_3}+\frac{\alpha_3\mu_0}
{|\Lambda_3|}=\mu-m_d-m_s
\>\eqno(10)$$
and, as a consequence, the corresponding meson wave functions of 
(4) are represented by (6) and (7). 
It is evident for employing the above relations we have to assign some values 
to quark masses and gauge coupling constant $g$. We take the current quark 
masses used in \cite{{Gon07a},{Gon07b},{Gon08},{Gon08a},{Gon08b}} and they are
$m_d=5\>\,{\rm MeV}$, $m_s=107.5\>\,{\rm MeV}$. 
Under the circumstances, the reduced mass $\mu_0$ of (5) will be equal to 
$m_dm_s/(m_d+m_s)$. As to the 
gauge coupling constant $g=\sqrt{4\pi\alpha_s}$, it should be noted that 
recently some attempts have been made to generalize the standard formula
for $\alpha_s=\alpha_s(Q^2)=12\pi/[(33-2n_f)\ln{(Q^2/\Lambda^2)}]$ ($n_f$ is 
number of quark flavours) holding true at the momentum transfer 
$\sqrt{Q^2}\to\infty$ 
to the whole interval $0\le \sqrt{Q^2}\le\infty$. If employing one such a 
generalization used in Refs. \cite{De1} which we have already discussed 
elsewhere 
(for more details see \cite{{Gon07a},{Gon07b},{Gon08},{Gon08a},{Gon08b}}) 
then (when fixing $\Lambda=0.234$ GeV, $n_f=3$) we obtain   
$g\approx5.290449085$ necessary for 
our further computations at the mass scale of neutral kaons.

\subsection{Electric form factor and the root-mean-square radius}
The relations (4), (6) and (7) allow us to compute an electric formfactor 
of a meson as a function of the square of momentum transfer $Q^2$ in the form 
(for more details see \cite{{Gon07a},{Gon07b},{Gon08},{Gon08a},{Gon08b}})
$$ f(Q^2)=\sum\limits_{j=1}^3f_j(Q^2)=$$
$$\sum\limits_{j=1}^3\frac{(2\beta_j)^{2\alpha_j+1}}{6\alpha_j}\cdot
\frac{\sin{[2\alpha_j\arctan{(\sqrt{|Q^2|}/(2\beta_j))]}}}
{\sqrt{|Q^2|}(4\beta_j^2-Q^2)^{\alpha_j}}\> \eqno(11) $$
which also entails the root-mean-square radius of the meson (quarkonium) 
in the form 
$$<r>=\sqrt{\sum\limits_{j=1}^3\frac{2\alpha^2_j+3\alpha_j+1}
{6\beta_j^2}}\eqno(12)$$
that is in essence a radius of confinement.

\subsection{Magnetic moment}
Also it is not complicated to show with the help (4), (6) and (7) that 
the magnetic moments of mesons (quarkonia) with the 
wave functions of (4) (at $l_j=0$) are equal to zero 
\cite{{Gon07a},{Gon07b},{Gon08},{Gon08a},{Gon08b}}, as should be according 
to experimental data \cite{pdg}. 

Though we can also evaluate the magnetic form factor $F(Q^2)$ of meson 
(quarkonium) which is also a function of $Q^2$ (see Refs. \cite{{Gon07a},{Gon07b}}) 
the latter will not be used in the given paper so we shall not dwell upon it.

\section{Estimates for parameters of SU(3)-gluonic field in 
neutral kaons}
\subsection{Basic equations and numerical results}
Now we are able to estimate parameters $a_j, b_j, B_j$ of the confining 
SU(3)-field (3) for neutral kaons within framework of our approach. 
In this situation, we should consider (10) and (12) the system of equations 
which should be solved compatibly if taking $\mu= 497.648$ MeV, 
$m_d= 5.0$ MeV, $m_s= 107.5$ MeV and $<r>\approx0.560$ fm in accordance 
with \cite{pdg}. While computing 
for distinctness we take all eigenvalues $\lambda_j$ of the Euclidean Dirac 
operator ${\mathcal D}_0$ on the unit 2-sphere ${\mathbb S}^2$ equal to 1. The 
results of numerical compatible solving of equations (10) and (12)
are adduced in Tables 1--2.

\begin{table*}
\caption{Gauge coupling constant, reduced mass $\mu_0$ and
parameters of the confining SU(3)-gluonic field for neutral kaons}
\label{tab:1}
%\begin{center}
\begin{tabular}{|l|l|l|l|l|l|l|l|l|}
\hline\noalign{\smallskip}
%\noalign{\hrule}\\ 
\small Particle & \small $ g$ & \small $\mu_0$ (\small MeV) & \small $a_1$ 
& \small $a_2$ & \small $b_1$ (\small GeV) & \small $b_2$ (\small GeV) 
& \small $B_1$ & \small $B_2$ \\
\noalign{\smallskip}\hline\noalign{\smallskip}
%\noalign{\hrule}\\
\scriptsize $K^0,\bar{K}^0$ -- $d\bar{s}$, $\bar{d}s$  
& \scriptsize 5.29045
& \scriptsize 4.77778
& \scriptsize 0.102484
& \scriptsize -0.198658
& \scriptsize 0.385250
& \scriptsize -0.130208
& \scriptsize  -0.360
& \scriptsize   -0.170 \\
%\noalign{\hrule}\\
\noalign{\smallskip}\hline
\end{tabular}
%\end{center}
\end{table*}

\begin{table*}
\caption{Theoretical and experimental mass and radius 
of neutral kaons}
\label{t.2}
\begin{center}
\begin{tabular}{|l|l|l|l|l|} 
\hline
\tiny Particle & \tiny Theoret. $\mu$ (MeV) &  \tiny Experim. $\mu$ (MeV) & 
\tiny Theoret. $<r>$ (fm)  & \tiny Experim. $<r>$ (fm)  \\
\hline
\scriptsize $K^0,\bar{K}^0$ -- $d\bar{s}$, $\bar{d}s$
& \scriptsize $\mu= m_d+m_s+\omega_j(0,0,1)= 497.648$
& \scriptsize 497.648
& \scriptsize 0.550510
& \scriptsize 0.560 \\
\hline
\end{tabular}
\end{center}
\end{table*}

\subsection{Consistency with the widths of 2-photon decays
$K^0,\bar{K}^0\to2\gamma$} 
Let us consider whether the estimates of previous subsection are consistent 
with the width of the electromagnetic 2-photon decays  
$K^0,\bar{K}^0\to2\gamma$. 
Actually kinematic analysis based on 
Lorentz- and gauge invariances gives rise to the following expression for 
the width $\Gamma$ of the electromagnetic decay $P\to2\gamma$ (where 
$P$ stands for any meson from 
$\pi^0$, $\eta$, $\eta^\prime$, $K^0,\bar{K}^0$, see, e.g., Ref. \cite{RF})
$$ \Gamma=\frac{1}{4}\pi\alpha_{em}^2g^2_{P\gamma\gamma}\mu^3 \eqno(13) $$
with the electromagnetic coupling constant $\alpha_{em}$=1/137.0359895 and 
the $P$-meson mass $\mu$ while the information about strong 
interaction of quarks in $P$-meson is encoded in a decay constant 
$g_{P\gamma\gamma}$. Making replacement $g_{P\gamma\gamma}=
f_P/\mu $ we can reduce (13) to the form 
$$ \Gamma=\frac{\pi\alpha_{em}^2\mu f_P^2}{4}\>. \eqno(14) $$
Now it should be noted that the only 
invariant which $f_P$ might depend on is $Q^2=\mu^2$, i. e. we should find 
such a function ${\mathcal F}(Q^2)$ for that ${\mathcal F}(Q^2=\mu^2)=f_P$ but  
${\mathcal F}(Q^2)$ cannot be computed by perturbative techniques. It is 
obvious from the physical point of view that ${\mathcal F}(Q^2)$ should be connected 
with the electromagnetic properties of $P$-meson. As we have seen in 
Section 3, there are at least two suitable functions for this aim -- electric 
and magnetic form factors. But there exist no experimental 
consequences related to a magnetic form factor at present whereas electric 
one to some extent determines, e. g., an effective size of meson (quarkonium) 
in the 
form $<r>$ of (12). It is reasonable, therefore, to take 
${\mathcal F}(Q^2=\mu^2)=Af(Q^2=\mu^2)$ with some constant $A$ and the electric form 
factor $f$ of (11) for the sought relation. In this situation, we obtain 
an additional equation imposed on parameters of the confining SU(3)-gluonic field 
in $P$-meson which has been used in Refs. \cite{{Gon07a},{Gon07b}} to 
estimate the mentioned parameters in $\pi^0$- and $\eta$-mesons. 
As a result, using (11) in the case of neutral kaons, we come from 
(14) to relation 
$$ \Gamma=\frac{\pi\alpha_{em}^2\mu}{4}
\left(A\sum\limits_{j=1}^3\frac{1}{6\alpha_jx_j}\cdot
\frac{\sin{(2\alpha_j\arctan{x_j})}}
{(1-x_j^2)^{\alpha_j}}\right)^2\approx$$
$$\cases{0.209\times10^{-10} \,{\rm eV},\, K^0_S-{\rm mode},\cr
0.696\times10^{-11} \,{\rm eV}, \,K^0_L-{\rm mode} \cr}\> \eqno(15) $$
with $x_j=\mu/(2\beta_j)$, $\mu=497.648$ MeV and we used widths 
$\Gamma_7\approx0.209\times10^{-10} \,{\rm eV}$, 
$\Gamma_{17}\approx0.696\times10^{-11} \,{\rm eV}$ for decays 
$K^0,\bar{K}^0\to2\gamma$, respectively, for $K^0_S$ - and $K^0_L$-modes 
following the notation from Ref. \cite{pdg}. In the circumstances, 
we can employ the results of Table 1 and compute the left-hand side of 
(15) which entails 
the corresponding values $A\approx0.3263\times10^{-7}$ and 
$A\approx0.1884\times10^{-7}$. Consequently, 
we draw the conclusion that parameters of the confining SU(3)-gluonic field 
in neutral kaons from Table 1 might be consistent with $\Gamma_7$ and 
$\Gamma_{17}$ while smallness of constants $A$ indicates the electromagnetic 
properties of neutral kaons to be inessential.

\section{Estimates of gluon concentrations, electric and magnetic colour field 
strengths}
Now let us recall that, according to Refs. \cite{{Gon052},{Gon06}}, one can 
confront the field (3) with the $T_{00}$-component (the volumetric energy 
density of the SU(3)-gluonic field) of the energy-momentum tensor (2) so that 
$$T_{00}\equiv T_{tt}=\frac{E^2+H^2}{2}=$$
$$\frac{1}{2}\left(\frac{a_1^2+
a_1a_2+a_2^2}{r^4}+\frac{b_1^2+b_1b_2+b_2^2}{r^2\sin^2{\vartheta}}\right)
\equiv\frac{{\mathcal A}}{r^4}+
\frac{{\mathcal B}}{r^2\sin^2{\vartheta}}\>\eqno(16)$$
with electric $E$ and magnetic $H$ colour field strengths and with 
real ${\mathcal A}>0$, ${\mathcal B}>0$. One can also introduce magnetic colour 
induction $B=(4\pi\times10^{-7} {\rm H/m})\,H$, where $H$ in A/m.  

To estimate the gluon concentrations
we can employ (16) and, taking the quantity
$\omega= \Gamma$, the full decay width of a meson, for 
the characteristic frequency of gluons we obtain
the sought characteristic concentration $n$ in the form
$$n=\frac{T_{00}}{\Gamma}\>, \eqno(17)$$
so we can rewrite (16) in the form 
$T_{00}=T_{00}^{\rm coul}+T_{00}^{\rm lin}$ conforming to the contributions 
from the Coulomb and linear parts of the
solution (3). This entails the corresponding split of $n$ from (17) as 
$n=n_{\rm coul} + n_{\rm lin}$. 

The parameters of Table 1 were employed when computing and for simplicity 
we put $\sin{\vartheta}=1$ in (16). There was also used the following 
present-day full decay widths of mesons under consideration \cite{pdg}: 
${\Gamma}=1/\tau$ with the life times $\tau= 
0.8953\times10^{-10}$ s ($K^0_S$-mode), $5.18\times10^{-8}$ s ($K^0_L$-mode), 
respectively, whereas the Bohr radius 
$a_0=0.529177249\cdot10^{5}\ {\rm fm}$ \cite{pdg}. 

Table 3 contains the numerical results for $n_{\rm coul}$, $n_{\rm lin}$, $n$, 
$E$, $H$, $B$ for the mesons under discussion.
\begin{table*}
\caption{Gluon concentrations, electric and magnetic colour field strengths in 
neutral kaons}
\label{t.3}
\begin{center}
\begin{tabular}{|lllllll|}
%\hline
%\noalign{\hrule}\\
\hline
\scriptsize $K^0_S$-mode: & \scriptsize 
$r_0=<r>= 0.550510 \ {\rm fm}$ & & &  & & \\
\hline
%\noalign{\hrule}\\
\tiny $r$ (fm)& \tiny $n_{\rm coul}$ $ ({\rm m}^{-3}) $ & \tiny $n_{\rm lin}$ 
$ ({\rm m}^{-3}) $& \tiny $n$ (${\rm m}^{-3}) $ & \tiny $E$ $({\rm V/m})$ 
& \tiny $H$ $({\rm A/m})$ & \tiny $B$ $({\rm T})$\\
\hline
%\noalign{\hrule}\\
\tiny $0.1r_0$ 
& \tiny $ 0.285346\times10^{65}$   
& \tiny $ 0.336488\times10^{63}$ 
& \tiny $ 0.288711\times10^{65}$ 
& \tiny $ 0.957080\times10^{24}$  
& \tiny $ 0.139807\times10^{22}$ 
& \tiny $ 0.175687\times10^{16}$\\
\hline
\tiny$r_0$ 
& \tiny$ 0.285346\times10^{61}$ 
& \tiny$ 0.336488\times10^{61}$ 
& \tiny$ 0.621834\times10^{61}$
& \tiny$ 0.957080\times10^{22}$  
& \tiny$ 0.139807\times10^{21}$ 
& \tiny$ 0.175687\times10^{15}$ \\
\hline
\tiny$1.0$ 
& \tiny$ 0.262079\times10^{60}$  
& \tiny$ 0.101976\times10^{61}$ 
& \tiny$ 0.128184\times10^{61}$ 
& \tiny$ 0.290054\times10^{22}$  
& \tiny$ 0.769654\times10^{20}$
& \tiny$ 0.967175\times10^{14}$ \\
\hline
\tiny$10r_0$ 
& \tiny$ 0.285346\times10^{57}$  
& \tiny$ 0.336488\times10^{59}$ 
& \tiny$ 0.339342\times10^{59}$ 
& \tiny$ 0.957080\times10^{20}$  
& \tiny$ 0.139807\times10^{20}$ 
& \tiny$ 0.175687\times10^{14}$ \\
\hline
\tiny$a_0$ 
& \tiny$ 0.334217\times10^{41}$  
& \tiny$ 0.364165\times10^{51}$ 
& \tiny$ 0.364165\times10^{51}$ 
& \tiny$ 0.103580\times10^{13}$ 
& \tiny$ 0.145443\times10^{16}$  
& \tiny$ 0.182770\times10^{10}$ \\
\hline
%\noalign{\hrule}\\
\scriptsize $K^0_L$-mode: & \scriptsize 
$r_0=<r>= 0.550510\ {\rm fm}$  &  & &  & & \\
\hline 
%\noalign{\hrule}\\
\tiny $r$ (fm)& \tiny $n_{\rm coul}$ $ ({\rm m}^{-3}) $ & \tiny $n_{\rm lin}$ 
$ ({\rm m}^{-3}) $& \tiny $n$ (${\rm m}^{-3}) $ & \tiny $E$ $({\rm V/m})$ 
& \tiny $H$ $({\rm A/m})$ & \tiny $B$ $({\rm T})$\\
\hline
%\noalign{\hrule}\\
\tiny $0.1r_0$ 
& \tiny $ 0.165095\times10^{68}$   
& \tiny $ 0.194684\times10^{66}$ 
& \tiny $ 0.167042\times10^{68}$ 
& \tiny $ 0.957080\times10^{24}$  
& \tiny $ 0.139807\times10^{22}$ 
& \tiny $ 0.175687\times10^{16}$ \\
\hline
\tiny$r_0$ 
& \tiny$ 0.165095\times10^{64}$ 
& \tiny$ 0.194684\times10^{64}$ 
& \tiny$ 0.359779\times10^{64}$
& \tiny$ 0.957080\times10^{22}$  
& \tiny$ 0.139807\times10^{21}$ 
& \tiny $ 0.175687\times10^{15}$ \\
\hline
\tiny$1.0$ 
& \tiny$ 0.151633\times10^{63}$  
& \tiny$ 0.590013\times10^{63}$ 
& \tiny$ 0.741646\times10^{63}$ 
& \tiny$ 0.290054\times10^{22}$  
& \tiny$ 0.769654\times10^{20}$  
& \tiny$ 0.967175\times10^{14}$ \\  
\hline
\tiny$10r_0$ 
& \tiny $ 0.165095\times10^{60}$  
& \tiny $ 0.194684\times10^{62}$ 
& \tiny $ 0.196335\times10^{62}$ 
& \tiny $ 0.957080\times10^{20}$  
& \tiny $ 0.139807\times10^{20}$ 
& \tiny $ 0.175687\times10^{14}$ \\
\hline
\tiny$a_0$ 
& \tiny $ 0.193370\times10^{44}$  
& \tiny $ 0.210697\times10^{54}$ 
& \tiny $ 0.210697\times10^{54}$ 
& \tiny $ 0.103580\times10^{13}$ 
& \tiny $ 0.145443\times10^{16}$ 
& \tiny $ 0.182770\times10^{10}$ \\ 
\hline
%\noalign{\hrule}\\
%\noalign{\hrule}\\
\end{tabular}
\end{center}
\end{table*}

\subsection{Concluding remarks}
 As is seen from Table 3, at the characteristic scales
of neutral kaons the gluon concentrations are huge and the corresponding 
fields (electric and magnetic colour ones) can be considered to be 
the classical ones with enormous strenghts. The part $n_{\rm coul}$ of gluon 
concentration $n$ connected with the Coulomb electric colour field is 
decreasing faster than $n_{\rm lin}$, the part of $n$ related to the linear 
magnetic colour field, and at large distances $n_{\rm lin}$ becomes dominant. 
It should be emphasized that in fact the gluon concentrations are much 
greater than the estimates given in Table 3 
because the latter are the estimates for maximal possible gluon frequencies, 
i.e. for maximal possible gluon impulses (under the concrete situation of 
neutral kaons). As was mentioned in section 1, 
the overwhelming majority of gluons between quarks is soft, i. e., with 
frequencies much less than $\Gamma=1/\tau$ with the life time $\tau$ for 
$K^0_S$ or $K^0_L$, so the corresponding concentrations are much greater 
than those in table 3. The given picture is in concordance with 
the one obtained 
in \cite{{Gon03},{Gon06},{Gon07a},{Gon07b},{Gon08a},{Gon08b}}. 
As a result, the confinement mechanism developed 
in \cite{{Gon01},{Gon051},{Gon052}}
and described early in section 1 is also confirmed by the considerations of 
the present paper. 

It should be noted, however, that our results are of a preliminary character 
which is readily apparent, for example, from the fact that the current quark masses 
(as well as the gauge coupling constant $g$) used in computation are known 
only within the certain limits, and we can expect similar limits for the 
magnitudes discussed in the paper so it is neccesary for further specification 
of the parameters for the confining SU(3)-gluonic field 
in neutral kaons which can be obtained, for instance, by calculating 
the widths of decays $K^0_S\to\pi^+\pi^-$ or $K^0_L\to3\pi^0$ 
with the help of wave functions discussed above 
and in \cite{{Gon07a},{Gon08b}}. We hope to continue analysing 
the given problems elsewhere. 

\section{Chiral limit for pseudoscalar nonet}
Having obtained estimates for neutral kaons in previous sections we can 
state that at the given moment we have such estimates for all the members of 
pseudoscalar nonet if taking into account the results of 
\cite{{Gon07a},{Gon07b},{Gon08},{Gon08b}}. Under the circumstances we can 
return to the 
chiral symmetry breaking problem in QCD whose possible resolution within the 
framework of the above confinement mechanism has been discussed in 
\cite{Gon08b}. As was mentioned in section 2, merits of case consists in that 
the Dirac equation in the field (3) possesses a nontrivial spectrum of bound 
states even for massless fermions [(see relation (9)]. As a result, 
mass of any meson remains nonzero in chiral limit when masses of quarks 
$m_q\to0$ and meson masses will only be expressed through the parameters of 
the confining SU(3)-gluonic field of (3). This purely gluonic residual mass of 
meson should be interpreted as a gluonic contribution to the meson mass.

Physically this gives us  
a possible approach to the problem of chiral symmetry breaking in QCD 
\cite{Gon08b}: in chirally symmetric world masses of mesons are fully 
determined by the confining SU(3)-gluonic field between (massless) quarks 
and are not equal to zero. Accordingly chiral symmetry is a sufficiently rough 
approximation holding true only when neglecting the mentioned SU(3)-gluonic 
field between quarks and no additional mechanism of the spontaneous chiral 
symmetry breaking connected to the so-called Goldstone bosons is required. 
Referring for more details to \cite{Gon08b}, we can here only say that, e.g., 
masses of mesons from pseudoscalar nonet have a purely 
gluonic contribution and we may be interested in what part of the meson masses  
is obligatory to that contribution. Therefore, let us employ the results 
of both \cite{{Gon07a},{Gon07b},{Gon08},{Gon08b}} and the present paper to estimate 
the mentioned contribution for all the members of pseudoscalar nonet. To pass 
on to obtaining the sought estimates all the necessary parameters gained in 
\cite{{Gon07a},{Gon07b},{Gon08},{Gon08b}} are gathered in table 4 where we took into 
account that in accordance with the standard quark 
model based on SU(3)-flavour symmetry (see, e.g., \cite{pdg}) 
$\pi^0=(\bar{u}u-\bar{d}d)/\sqrt{2}$ is a superposition of two quarkonia while 
$\eta=(2\bar{s}s-\bar{u}u-\bar{d}d)/\sqrt{6}$ and 
$\eta^\prime=(\bar{u}u+\bar{d}d+\bar{s}s)/\sqrt{3}$ are the superpositions of 
three quarkonia so we have, respectively, two or three sets of parameters 
$a_j$, $b_j$, $B_j$ for the corresponding particles. 

\begin{table*}
\caption{Gauge coupling constant, reduced mass $\mu_0$ and
parameters of the confining SU(3)-gluonic field for pseudoscalar nonet.}
\label{t.4}
\begin{center}
\begin{tabular}{|l|l|l|l|l|l|l|l|l|}
\hline
%\noalign{\hrule}\\ 
\small Particle & \small $ g$ & \small $\mu_0$ (\small MeV) & \small $a_1$ 
& \small $a_2$ & \small $b_1$ (\small GeV) & \small $b_2$ (\small GeV) 
& \small $B_1$ & \small $B_2$ \\
\hline
%\noalign{\hrule}\\
\scriptsize $\pi^{0}$ -- $\bar{u}u$  
& \scriptsize 6.10148 
& \scriptsize 1.125 
& \scriptsize -0.0434737 
& \scriptsize -0.00680835 
& \scriptsize 0.0848234 
& \scriptsize 0.0433136 
& \scriptsize 0.01 
& \scriptsize  -0.150 \\
\hline
\scriptsize $\pi^{0}$ -- $\bar{d}d$  
& \scriptsize 6.10148 
& \scriptsize 2.50 
& \scriptsize -0.0606679 
& \scriptsize 0.0251427 & 
\scriptsize 0.0956303 
& \scriptsize 0.0648174 
& \scriptsize 0.1250 
& \scriptsize  -0.2450 \\
%\noalign{\hrule}\\
\hline
\scriptsize $\pi^\pm$ -- $u\bar{d}$, $\bar{u}d$  
& \scriptsize 6.09131
& \scriptsize 1.55172
& \scriptsize 0.0473002
& \scriptsize 0.0118497
& \scriptsize 0.178915 
& \scriptsize -0.119290
& \scriptsize -0.230
& \scriptsize  0.230 \\
\hline
\scriptsize $K^\pm$ -- $u\bar{s}$, $\bar{u}s$  
& \scriptsize 5.30121
& \scriptsize 2.20387
& \scriptsize 0.167182
& \scriptsize -0.0557501
& \scriptsize 0.120150
& \scriptsize 0.131046
& \scriptsize -0.900
& \scriptsize  0.290 \\
\hline
\scriptsize $K^0,\bar{K}^0$ -- $d\bar{s}$, $\bar{d}s$  
& \scriptsize 5.29045
& \scriptsize 4.77778
& \scriptsize 0.102484
& \scriptsize -0.198658
& \scriptsize 0.385250
& \scriptsize -0.130208
& \scriptsize  -0.360
& \scriptsize   -0.170 \\
%\noalign{\hrule}\\
\hline
\scriptsize $\eta$ -- $\bar{u}u$  
& \scriptsize 5.14836 
& \scriptsize 1.125 
& \scriptsize -0.0328122
& \scriptsize 0.179728 
& \scriptsize 0.194979 
& \scriptsize 0.119737 
& \scriptsize 0.255
& \scriptsize  -0.010 \\
\hline
\scriptsize $\eta$ -- $\bar{d}d$  
& \scriptsize 5.14836 
& \scriptsize 2.50 
& \scriptsize 0.147640
& \scriptsize -0.178707 
& \scriptsize 0.305728 
& \scriptsize -0.119050 
& \scriptsize -0.240
& \scriptsize  -0.010 \\  
\hline    
\scriptsize $\eta$ -- $\bar{s}s$  
& \scriptsize 5.14836 
& \scriptsize 53.75  
& \scriptsize -0.0141391 
& \scriptsize -0.0806779 
& \scriptsize 0.252975 
& \scriptsize -0.339250 
& \scriptsize 0.260 
& \scriptsize -0.310 \\ 
\hline
\scriptsize $\eta^\prime$ -- $\bar{u}u$  
& \scriptsize 3.91476 
& \scriptsize 1.125 
& \scriptsize 0.218474 
& \scriptsize -0.394718 
& \scriptsize 0.618419 
& \scriptsize -0.280807 
& \scriptsize -0.300 
& \scriptsize  -0.200 \\
\hline
\scriptsize $\eta^\prime$ -- $\bar{d}d$  
& \scriptsize 3.91476 
& \scriptsize 2.50 
& \scriptsize 0.351384 
& \scriptsize -0.130858 
& \scriptsize 0.278983 
& \scriptsize 0.285548 
& \scriptsize -0.150 
& \scriptsize  0.410 \\
\hline
\scriptsize $\eta^\prime$ -- $\bar{s}s$  
& \scriptsize 3.91476 
& \scriptsize 53.75 
& \scriptsize 0.123645 
& \scriptsize 0.124633 
& \scriptsize -0.226875 
& \scriptsize 0.588802 
& \scriptsize 0.410 
& \scriptsize  -0.160 \\
%\noalign{\hrule}\\
\hline
\end{tabular}
\end{center}
\end{table*}

The current quark masses were taken with 
the same values as in \cite{{Gon07a},{Gon07b},{Gon08},{Gon08b}}, i.e.,  
$m_u=2.25\>\,{\rm MeV}$, $m_d=5\>\,{\rm MeV}$, $m_s=107.5\>\,{\rm MeV}$. 

Now we can note that according to our approach the mass of any meson from 
table 4 is given by relation [cf. (10)] 
$$\mu=m_{q_1}+m_{q_2}+\frac{g^2a_1b_1}{\lambda_1-gB_1}+
\mu_0\frac{\sqrt{(\lambda_1-gB_1)^2-g^2a_1^2}}
{|\lambda_1-gB_1|}= $$
$$m_{q_1}+m_{q_2}+\frac{g^2a_2b_2}{\lambda_2-gB_2}+
\mu_0\frac{\sqrt{(\lambda_2-gB_2)^2-g^2a_2^2}}
{|\lambda_2-gB_2|}=$$
$$=m_{q_1}+m_{q_2}+\frac{g^2(a_1+a_2)(b_1+b_2)}{\lambda_3+g(B_1+B_2)}+$$
$$\mu_0\frac{\sqrt{[\lambda_3+g(B_1+B_2)]^2-g^2(a_1+a_2)^2}}{|\lambda_3+
g(B_1+B_2)|}\>,$$
$$ \mu_0=\frac{m_{q_1}m_{q_2}}{m_{q_1}+m_{q_2}},\eqno(18)$$
and the eigenvalues of the Euclidean Dirac operator on a unit sphere
$\lambda_j=-1$ for $\pi^0$ and $\lambda_j=1$ for the rest of 
particles. 

In chiral limit $m_{q_1}, m_{q_2}\to0$ 
we obtain 
$$(\mu)_{chiral}\approx\frac{g^2a_1b_1}{\lambda_1-gB_1}\approx
\frac{g^2a_2b_2}{\lambda_2-gB_2}\approx$$
$$\frac{g^2(a_1+a_2)(b_1+b_2)}{\lambda_3+g(B_1+B_2)}\ne0
\>.\eqno(19)$$
We can see that in 
chiral limit the meson masses are completely determined only 
by the parameters $a_j, b_j, B_j$ of SU(3)-gluonic field between quarks, i.e. 
by interaction between quarks, and those masses have the purely gluonic nature. 
So one can use the parameters $g, a_j, b_j, B_j$ adduced in table 4 
to compute $(\mu)_{chiral}$ which in fact represents the sought gluonic 
contribution to the meson masses. The results of computation are gathered in 
table 5. 

\begin{table*}
\caption{Theoretical, experimental and chiral meson masses}
\label{t.5}
\begin{center}
\begin{tabular}{|l|l|l|l|l|} 
\hline
\tiny Particle & \tiny Theoretical (MeV) &  \tiny Experimental (MeV) & 
\tiny Chiral (MeV)  & \tiny Gluonic contribution (\%)  \\
\hline
\scriptsize $\pi^{0}$ -- $\bar{u}u$
& \scriptsize $\mu= 2m_u+\omega_j(0,0,-1)= 134.976$
& \scriptsize 134.976
& \scriptsize 129.495
& \scriptsize 96.0 \\
\hline
\scriptsize $\pi^{0}$ -- $\bar{d}d$
& \scriptsize $\mu= 2m_d+\omega_j(0,0,-1)= 134.976$
& \scriptsize 134.976
& \scriptsize 122.5
& \scriptsize 90.8 \\
\hline
\scriptsize $\pi^\pm$ -- $u\bar{d}$, $\bar{u}d$
& \scriptsize $\mu= m_u+m_d+\omega_j(0,0,1)= 139.570$
& \scriptsize 139.56995
& \scriptsize 130.8 
& \scriptsize 93.7 \\
\hline
\scriptsize $K^\pm$ -- $u\bar{s}$, $\bar{u}s$
& \scriptsize $\mu= m_u+m_s+\omega_j(0,0,1)= 493.677$
& \scriptsize 493.677
& \scriptsize 382.0
& \scriptsize 77.4 \\
\hline
\scriptsize $K^0,\bar{K}^0$ -- $d\bar{s}$, $\bar{d}s$
& \scriptsize $\mu= m_d+m_s+\omega_j(0,0,1)= 497.648$
& \scriptsize 497.648
& \scriptsize 380.569
& \scriptsize  76.6 \\
\hline
\scriptsize $\eta$ -- $\bar{u}u$ 
& \scriptsize $\mu= 2m_u+\omega_j(0,0,1)= 547.51$
& \scriptsize 547.51
& \scriptsize 542.0
& \scriptsize 99.0  \\
\hline
\scriptsize $\eta$ -- $\bar{d}d$ 
& \scriptsize $\mu= 2m_d+\omega_j(0,0,1)= 547.51$
& \scriptsize 547.51
& \scriptsize 535.4
& \scriptsize 97.8  \\
\hline
\scriptsize $\eta$ -- $\bar{s}s$ 
& \scriptsize $\mu= 2m_s+\omega_j(0,0,1)= 547.51$
& \scriptsize 547.51
& \scriptsize 280.01
& \scriptsize 51.1  \\
\hline
\scriptsize $\eta^\prime$ -- $\bar{u}u$ 
& \scriptsize $\mu= 2m_u+\omega_j(0,0,1)= 957.78$
& \scriptsize 957.78
& \scriptsize 952.5
& \scriptsize 99.4  \\
\hline
\scriptsize $\eta^\prime$ -- $\bar{d}d$ 
& \scriptsize $\mu= 2m_d+\omega_j(0,0,1)= 957.78$
& \scriptsize 957.78
& \scriptsize 946.0
& \scriptsize 98.8  \\
\hline
\scriptsize $\eta^\prime$ -- $\bar{s}s$ 
& \scriptsize $\mu= 2m_s+\omega_j(0,0,1)= 957.78$
& \scriptsize 957.78
& \scriptsize 691.5
& \scriptsize 72.2  \\
\hline
\end{tabular}
\end{center}
\end{table*}

One can add to the results of table 5 that we could also calculate, e.g., 
the root-mean-square radii (12) of particles under consideration in chiral 
limit which are well defined in this limit as well. For example, 
$<r>_{chiral}\approx0.673069$ fm or 0.543223 fm, accordingly, for charged 
pions and kaons \cite{Gon08b}. I.e., those values only slightly differ 
from the present-day experimental values 0.672 fm or 0.560 fm \cite{pdg}. 
The same remark also holds true for the decay 
constants for leptonic decays $f_P$ ($P$ stands for charged pions and kaons, 
see \cite{Gon08b} for more details). So, even in chirally symmetric world, e.g., 
the charged pions and kaons would have nonzero masses, the root-mean-square 
radii and decay constants $f_P$ for leptonic decays and all of those quantities would be 
determined only by SU(3)-gluonic interaction between massless quarks, i.e. 
they would have a purely gluonic nature. 
Moreover, since gluons are verily relativistic 
particles then the most part of masses for mesons of pseudoscalar
nonet is conditioned by relativistic effects, as is seen from table 5. Further 
discussion of the proposed chiral symmetry breaking mechanism can be found 
in \cite{Gon08b}. 

\section{A possible relation with a phenomenological string-like picture of 
quark confinement}

\subsection{The confining potential and string tension}
The results adduced in section 5 allow us to shed some light on one more 
problem which has been touched upon in \cite{{Gon06},{Gon08b}}. 
As is known, for a long time up to now 
there exists the so-called string-like picture of quark
confinement but only at qualitative phenomenological level (see, e. g., 
Ref. \cite{Per}). Up to now, however, it is unknown how such a 
picture might be warranted from the point of view of QCD. Let us in short 
outline as our results for pseudoscalar nonet (based on and derived from 
QCD-Lagrangian directly) naturally lead to possible justification of the 
mentioned contruction. Thereto 
we note that one can calculate energy ${\mathcal E}$ of gluon condensate conforming 
to solution (3) in a volume $V$ through relation 
${\mathcal E}=\int_VT_{00}r^2\sin{\vartheta}dr d\vartheta d\varphi\>$ with $T_{00}$ 
of (16) but one should take into account that classical $T_{00}$ has a 
singularity along $z$-axis ($\vartheta=0,\pi$) and we have to introduce some 
angle $\vartheta_0$ so $\vartheta_0\leq\vartheta\leq\pi-\vartheta_0$.  
As well as in Ref. \cite{Gon051}, we may consider $\vartheta_0$ to be 
a parameter determining some cone $\vartheta=\vartheta_0$ so the quark  
emits gluons outside of the cone. Now if there are two quarks $Q_1, Q_2$ and 
each of them emits gluons outside 
of its own cone $\vartheta=\vartheta_{1,2}$ (see Figs. 1, 2) then we have 
soft gluons (as mentioned in section 1) in regions I, II and between quarks.  

Accordingly, we shall have some region $V$ with gluon condensate between quarks 
$Q_1, Q_2$ and its vertical projection is shown in Fig. 1. Another projection 
of $V$ onto a plane perpendicular to the one of Fig. 1 is sketched out in 
Fig. 2.  
\begin{figure*}
\vspace{0cm}
\caption{Vertical projection of region with the gluon condensate energy 
between quarks.}
\end{figure*}

\begin{figure*}
\vspace{0cm}
\caption{Horizontal projection of region with the gluon condensate energy 
between quarks.}
\end{figure*}

Then, as is clear from Fig. 1, for distance $R$ between quarks we have 
$R=R_1\sin{\vartheta_1}+R_2\sin{\vartheta_2}$ and gluonic energy between
quarks will be equal to

$${\mathcal V}(R)=\int_VT_{00}r^2\sin{\vartheta}dr d\vartheta d\varphi=$$
$$\int_{r_1}^{R_1}\int^{\pi-\vartheta_1}_{\vartheta_1}
\int_{-\varphi_1}^{\varphi_1}
\left(\frac{{\mathcal A}}{r^2}+\frac{{\mathcal B}}{\sin^2{\vartheta}}\right)
\sin{\vartheta}dr d\vartheta d\varphi+$$
$$\int_{r_2}^{R_2}\int^{\pi-\vartheta_2}_{\vartheta_2}
\int_{-\varphi_2}^{\varphi_2}
\left(\frac{{\mathcal A}}{r^2}+\frac{{\mathcal B}}{\sin^2{\vartheta}}\right)
\sin{\vartheta}dr d\vartheta d\varphi
\>\eqno(20)$$
with constants ${\mathcal A}$, ${\mathcal B}$ defined in (16). 

To clarify a physical meaning of the quantities $r_{1,2}$ in Figs. 1, 2, 
let us recall an analogy with classical 
electrodynamics where is well known (see, e. g., \cite{LL} and 
Subsection 2.2) that the notion 
of classical electromagnetic field (a photon condensate) generated by a 
charged particle is applicable only at distances much greater than the Compton 
wavelength 
$\lambda_c=1/m$ for the given particle with mass $m$. Within the QCD framework 
the parameter $\Lambda_{QCD}$ plays a similar part (see, e.g., 
Ref. \cite{{pdg},{pi0}}). 
Namely, the notion of classical SU(3)-gluonic field ( a gluon condensate) is 
not applicable at the distances much 
less than $1/\Lambda_{QCD}$. In accordance with subsection 2.5 we took 
$\Lambda_{QCD}=\Lambda=0.234$ GeV which entails $1/\Lambda\sim$ 0.8433 fm 
so one may consider 
$r_{1,2}\sim$ 0.1$<r>$ with the root-mean-square radius $<r>$ for a meson.

Under the circumstances, performing a simple integration in (20) with 
employing the relations $\int d\vartheta/\sin{\vartheta}=\ln\tan{\vartheta/2}$, 
$\tan{\vartheta/2}=\sin{\vartheta}/(1+ \cos{\vartheta})=
(1-\cos{\vartheta})/\sin{\vartheta}$, we shall without going into details 
(see also Ref. \cite{Gon051}) obtain  

$${\mathcal V}(R_1,R_2)={\mathcal V}_0-\sum_{i=1}^2
\frac{{4\varphi_i\mathcal A}\cos{\vartheta_i}}{R_i}+\sum_{i=1}^2 
{2\varphi_i\mathcal B}R_i\ln\frac{1+                 
\cos{\vartheta_i}}{1-\cos{\vartheta_i}}, \eqno(21)$$
where  $${\mathcal V}_0=\sum_{i=1}^2{\mathcal V}_{0i}=$$
$$\sum_{i=1}^2\left(\frac{{4\varphi_i\mathcal A}\cos{\vartheta_i}}{r_i}-
{2\varphi_i\mathcal B}r_i\ln\frac{1+\cos{\vartheta_i}}
{1-\cos{\vartheta_i}}\right).$$ 

For the sake of simplicity let us put $R_1=R_2$, 
$\vartheta_1=\vartheta_2=\vartheta_0$, $\varphi_1=\varphi_2=\varphi_0$. Then 
$R_1=R_2=R/(2\sin{\vartheta_0})$ and from (21) it follows
$${\mathcal V}(R)= {\mathcal V}_0+\frac{a}{R}+kR \eqno(22)$$
with $a=-8\varphi_0{\mathcal A}\sin{2\vartheta_0}$, 
$k=2\varphi_0\frac{\mathcal B}{\sin{\vartheta_0}}
\ln{\frac{1+\cos{\vartheta_0}}{1-\cos{\vartheta_0}}}$. 

We recognize the modeling confining potential in (22) which is often used 
when applying to meson and heavy quarkonia physics (see, e.g., \cite{Bra05}). 
We can, however, see that phenomenological parameters 
$a, k, {\mathcal V}_0$ of potential (22) are expressed through more fundamental 
parameters $a_j$, $b_j$ connected with the unique exact solution (3) of 
Yang-Mills equations describing confinement. One can notice that the quantity 
$k$ (string tension) is usually related to the so-called Regge slope 
$\alpha^\prime=1/(2\pi k)$ and in many if not all of the papers using 
potential approach it is accepted $k\approx 0.18$ GeV$^2$ 
(see, e. g., \cite{Bra05}).  
\subsection{Estimates of $\vartheta_0$, $\varphi_0$ for pseudoscalar
nonet}
Under the situation we have the equation
$$k=2\varphi_0\frac{\mathcal B}{\sin{\vartheta_0}}
\ln{\frac{1+\cos{\vartheta_0}}{1-\cos{\vartheta_0}}}\approx0.18 
\>\rm GeV^2\>\eqno(23)$$ 
with ${\mathcal B}=(b_1^2+b_1b_2+b_2^2)/2$, 
so let us employ (23) to estimate $\vartheta_0$, $\varphi_0$ if using the 
parameters adduced in table 4 for pseudoscalar nonet and also for the 
ground state of toponium $\eta_t$ for that we use the 
parametrization from \cite{Gon08a} with the values 
$a_1= 0.361253$, $a_2= 0.339442$, $b_1= 48.9402$ GeV, $b_2= 76.7974$ GeV for 
the parameters of solution (3). Results of computations are presented in 
table 6. 

\begin{table*}
\caption{Angular parameters determining the gluon condensate between quarks 
for pseudoscalar nonet and toponium ground state}
\label{t.6}
\begin{center}
\begin{tabular}{|l|l|l|}
\hline
%\noalign{\hrule}\\ 
\small Particle 
& \small $\vartheta_0 $ 
& \small $\varphi_0$ \\ 
\hline
\scriptsize $\pi^{0}$ -- $\bar{u}u$
& \scriptsize $10^\circ$ 
& \scriptsize $28.84^\circ$ \\
\hline
& \scriptsize $30^\circ$ 
& \scriptsize $153.6^\circ$ \\
\hline
\scriptsize $\pi^{0}$ -- $\bar{d}d$
& \scriptsize $10^\circ$ 
& \scriptsize $18.8^\circ$ \\
\hline
& \scriptsize $30^\circ$ 
& \scriptsize $100.2^\circ$ \\
\hline
%\noalign{\hrule}\\
\scriptsize $\pi^{\pm}$ -- $u\bar{d}$, $\bar{u}d$ 
& \scriptsize $30^\circ$ 
& \scriptsize $78.63^\circ$ \\
\hline
& \scriptsize $45^\circ$
& \scriptsize $166.16^\circ$ \\
\hline
\scriptsize $K^{\pm}$ -- $u\bar{s}$, $\bar{u}s$ 
& \scriptsize $45^\circ$
& \scriptsize $87.36^\circ$ \\
\hline
& \scriptsize $60^\circ$
& \scriptsize $171.68^\circ$ \\
\hline
\scriptsize $K^0,\bar{K}^0$ -- $d\bar{s}$, $\bar{d}s$
& \scriptsize $60^\circ$
& \scriptsize $70.6^\circ$ \\
\hline
& \scriptsize $70^\circ$
& \scriptsize $118.0^\circ$ \\
\hline
\scriptsize $\eta$ -- $\bar{u}u$
& \scriptsize $45^\circ$
& \scriptsize $54.7^\circ$ \\
\hline
& \scriptsize $60^\circ$
& \scriptsize $107.4^\circ$ \\
\hline
\scriptsize $\eta$ -- $\bar{d}d$
& \scriptsize $45^\circ$
& \scriptsize $58.0^\circ$ \\
\hline
& \scriptsize $60^\circ$
& \scriptsize $114.1^\circ$ \\
\hline
\scriptsize $\eta$ -- $\bar{s}s$
& \scriptsize $60^\circ$
& \scriptsize $87.2^\circ$ \\
\hline
& \scriptsize $70^\circ$
& \scriptsize $145.8^\circ$ \\
\hline
\scriptsize $\eta^\prime$ -- $\bar{u}u$
& \scriptsize $70^\circ$
& \scriptsize $47.3^\circ$ \\
\hline
& \scriptsize $80^\circ$
& \scriptsize $100.6^\circ$ \\
\hline
\scriptsize $\eta^\prime$ -- $\bar{d}d$
& \scriptsize $70^\circ$
& \scriptsize $56.9^\circ$ \\
\hline
& \scriptsize $80^\circ$
& \scriptsize $121.1^\circ$ \\
\hline
\scriptsize $\eta^\prime$ -- $\bar{s}s$
& \scriptsize $70^\circ$
& \scriptsize $75.5^\circ$ \\
\hline
& \scriptsize $80^\circ$
& \scriptsize $160.8^\circ$ \\
\hline
\scriptsize $\eta_t$ -- $\bar{t}t$ 
& \scriptsize $60^\circ$
& \scriptsize $(0.675\times10^{-3})^\circ$ \\
\hline
& \scriptsize $80^\circ$
& \scriptsize $(0.240\times10^{-2})^\circ$ \\
\hline
& \scriptsize $88^\circ$
& \scriptsize $(0.123\times10^{-1})^\circ$ \\
\hline         
%\noalign{\hrule}\\
\end{tabular}
\end{center}
\end{table*}        
If taking into account that only the values of $\vartheta_0$, $\varphi_0$ 
between 0 and $90^\circ$ are of physical meaning and, according to Figs. 1, 2, 
the gluon configuration between quarks will be similar to a string-like one 
under the condition $\vartheta_0\to\pi/2$,  $\varphi_0\to0$, then we can see 
from table 6 that the characteristic transverse sizes 
$D_{1,2}$ of the gluon condensate between quarks in fact tend to zero only 
in the case of heavy quarks, i.e., only for heavy quarks the gluon 
configuration between them might practically transform into a string. As a result, 
there arises the string-like picture of quark confinement but the latter seems 
to be warranted enough only for heavy quarks. It should be emphasized that 
string tension $k$ of (23) is determined just by parameters $b_{1,2}$ of linear 
magnetic colour field from solution (3) which indirectly confirms 
the dominant role of the mentioned field for confinement.

We cannot, however, speak about potential ${\mathcal V}(R)$ of (22) as describing 
some gluon configuration between quarks. 
It would be possible if the mentioned potential were a solution of Yang-Mills 
equations directly derived from QCD-Lagrangian since, from the QCD-point of 
view, any gluonic field should be a solution of Yang-Mills equations (as well 
as any electromagnetic field is by definition always a solution of Maxwell 
equations). 

In reality, as was shown in 
Refs. \cite{{Gon051},{Gon052}} (see also Appendix C), potential of form (22) 
cannot be a solution 
of the Yang-Mills equations if simultaneously $a\ne0, k\ne0$. Therefore, 
it is impossible to obtain compatible solutions of the 
Yang-Mills-Dirac (Pauli, Schr{\"o}dinger) system when inserting potential of form 
(22) into Dirac (Pauli, Schr{\"o}dinger) equation. So, we draw the conclusion 
(mentioned as far back as in Refs. \cite{Gon03} and elaborated more in detail 
in Ref. \cite{Gon08a}) that the potential approach 
seems to be inconsistent: it is not based on compatible 
nonperturbative solutions for the Dirac-Yang-Mills system derived from 
QCD-Lagrangian in contrast to our confinement mechanism. Actually potential 
approach for heavy quarkonia has been historically modeled on positronium 
theory. In the latter case, however, one uses the {\em unique} modulo square 
integrable solutions of Dirac (Schr{\"o}dinger) 
equation in the Coulomb field [condensate of huge number of (virtual) photons], 
i. e., one employs the {\em unique} compatible nonperturbative solutions of the 
Maxwell-Dirac (Schr{\"o}dinger) system directly derived from QED-Lagrangian to 
describe positronium (or hydrogen atom) spectrum. 

To summarize, from the point of view of our approach both potential 
and string-like pictures of confinement arise only as some {\em effective} 
models derived in a certain way from the more fundamental theory based on 
exact solution (3) of SU(3)-Yang-Mills equations. This conlusion is in 
concordance with the ones obtained in \cite{{Gon06},{Gon08b}}. 

\section{Problem of masses in particle physics}
\subsection{Preliminaries}
As is known \cite{pdg}, the generally accepted standard model with one 
Higgs doublet asserts that the masses of fundamental fermions (quarks and 
leptons) are acquired through the Higgs mechanism so for their   
masses $m_i$ we obtain (without taking mixings into account) 
$m_i=f_{i}v/\sqrt{2}$, where the vacuum Higgs condensate $v\approx246$ GeV and 
$i$ stands for quark and lepton flavours. But little is known about the coupling 
constants $f_{i}$ and much may be elucidated 
only with discovering Higgs bosons. The same holds true for the gauge bosons 
$W^{\pm}$, $Z$ where masses $m_W=ev/(2\sin{\theta_W})$, $m_Z=ev/(\sin{2\theta_W})$ with 
the so-called weak angle $\theta_W$ so that 
$\sin^2{\theta_W}\approx0.23$ and $e$ is the elementary electric charge. If 
taking into account that the mass of Higgs boson $m_H=\lambda v$ with 
a self-interaction constant $\lambda$ then it is clear that masses of all the 
abovementioned particles are proportional to $m_H$, and, consequently, the 
discovery of Higgs boson will not completely resolve the puzzle of origin 
of masses in particle physics -- the question will remain where the mass 
$m_H$ comes from not speaking already about the nature of the above 
miscellaneous constants $f_i$ and $\lambda$. 

At present, to our mind, one can single out two most promising approaches to 
a possible resolution for the mentioned problems: technicolour theories and 
preon models. Under the circumstances let us shortly outline how both these 
directions might be estimated from the point of view of our confinement 
mechanism and the chiral symmetry breaking one based on the latter  
and discussed above and in \cite{Gon08b}. 
\subsection{Technicolour theories}
Referring for more details concerning those models to both early references 
\cite{T1} and modern status of them (see, e.g., \cite{T2}) let us note the 
following. The main idea of acquiring masses, e.g., for $W^{\pm}$ and $Z$ 
bosons, consists in that a new set of the so-called techniquarks is postulated 
at the energy scale of order 1 TeV which interact with each other through 
the technigluons and it makes the massless techipions exist as Goldstone bosons. 
The latter give masses to $W^{\pm}$ and $Z$ after spontaneous symmetry breaking. 
It should be noted, however, those massless technipions appear as a result of 
violating chiral symmetry connected with technicolour QCD on the analogy with 
chiral symmetry breaking in usual QCD. But, as we have discussed in 
\cite{Gon08b} and in section 5, the hypothetical mechanism for chiral symmetry 
breaking in standard QCD with appearance of Goldstone bosons (pions) seems to 
fail because of pions can never be massless inasmuch as they have nonzero masses 
even in chirally symmetric world due to gluons. The same will also perfectly 
hold true for technipions which would always have nonzero masses due to 
technigluons since technicolour QCD should manifest the confinement mechanism 
similar to our one in usual QCD. Therefore, technicolour theories look rather 
doubtful from the point of view of our confinement mechanism. 
\subsection{Preon models}
Another cardinal approach to the problem of masses is connected with the 
preon models (see, e.g., \cite{Pr} and references therein). Under this approch 
quarks, leptons and gauge vector bosons are suggested to be composed of stable 
spin-1/2 preons, for example, existing in three flavours and being combined 
according to simple rules. The main theoretical objection to preon theories is 
the mass paradox which arises by virtue of the Heisenberg's uncertainty 
principle. Scattering experiments have shown \cite{Per} that quarks and 
leptons are point-like up to the scales of order $10^{-3}$ fm which corresponds to 
a preon mass of order 197 GeV (due to the uncertainty 
principle) if the preon is confined to a box of 
such a size, i.e. its mass will approximately be $0.4\times10^{5}$ times greater 
than, e.g., that of $d$-quark. 
Thus, the preon models are faced with a mass paradox: how could quarks or electrons 
be made of smaller particles that would have masses of many orders of magnitude 
greater than the fundamental fermion masses? The paradox might be resolved by 
the rather dubious postulate about a large binding force between preons 
cancelling their mass-energies. Our confinement mechanism points out the more 
physically acceptable way of overcoming these obstacles. If the interaction
among preons is decribed by a QCD-like theory based on, e.g., SU(N)-group with 
$N\ge2$ then, according to our results \cite{Gon051,Gon052}, such theories should 
also manifest confinement to generate masses decribed by relations similar 
to (5) and (9). This signifies that preons might possess small masses 
or be just massless and, as a result, mass paradox would be removed. 

\section{Concluding remarks}
          The results of present paper as well as the ones of 
\cite{{Gon03},{Gon06},{Gon07a},{Gon07b},{Gon08},{Gon08a},{Gon08b}} allow one 
to speak about the fact that the 
confinement mechanism elaborated in \cite{{Gon01},{Gon051},{Gon052}} 
gives new possibilities for considering many old problems of hadronic 
(meson) physics (such as nonperturbative computation of decay constants, masses 
and radii of mesons, chiral symmetry breaking and so forth) from the first 
principles of QCD immediately appealing 
to the quark and gluonic degrees of freedom. This is possible because the 
given mechanism is based on the unique family of compatible 
nonperturbative solutions for the Dirac-Yang-Mills system directly derived from 
QCD-Lagrangian and, as a result, the approach is itself nonperturbative, 
relativistic from the outset, admits self-consistent nonrelativistic limit 
and may be employed for any meson (quarkonium). Under the circumstances the 
words {\em quark and gluonic degrees of freedom} make exact sense: gluons come 
forward in the form of bosonic condensate described by parameters $a_j$, 
$b_j$, $B_j$ from the unique exact solution (3) of the Yang-Mills equations 
while quarks are represented by their current masses $m_q$. Though nature of 
the latter is not yet totally understandable (see section 7) but the confinement 
mechanism under discussion indicates a possible way of overcoming this puzzle - 
quarks might be composed from just massless preons whose interaction would 
be described by a gauge theory with confinement mechanism similar to that under 
discussion and the current quark masses might be generated along the lines 
discussed in sections 5 and 7.

The given paper to a great degree summarizes studying nonet of light 
pseudoscalar mesons realized in 
\cite{{Gon06},{Gon07a},{Gon07b},{Gon08},{Gon08b}} 
within the framework of our approach and we can ascertain the fact that, on the 
whole, this nonet can be described from the united point of view of our 
confinement mechanism. In line with the above, obviously, one 
should now pass on to vector mesons ($\rho$, $\phi$, $\omega$...) and also to 
the light scalar mesons whose 
nature has been controversial over 30 years \cite{Close02}. As is clear from 
Section 2 (see also Appendices A, B), there exists a large number of 
relativistic bound states in the confining SU(3)-gluonic field (3) so all the mentioned 
mesons can probably correspond to some of those states and be described by 
their own sets of parameters $a_j$, $b_j$, $B_j$ 
of solution (3). More important task is, however, to explore  
possible ways to extend the approach over baryons, in particular, 
over nucleons. In this situation we shall have to deal with a relativistic 3-body 
problem as follows from SQM. It is clear, however, that confinement mechanism 
under discussion (which in essence decribes the relativistic 2-body problem) 
should also occupy a fitting place in the 3-body constructions. We hope to 
develop the given direction elsewhere. 

\section*{Appendix A}
We here represent some results about eigenspinors of the Euclidean Dirac 
operator on two-sphere ${\Bbb S}^2$ employed in the main part of the paper. 

When separating variables in the Dirac equation (4) there naturally 
arises the Euclidean Dirac operator ${\cal D}_0$ on the unit two-dimensional 
sphere ${\Bbb S}^2$ and we should know its eigenvalues with the corresponding 
eigenspinors. Such a problem also arises in the black hole theory while 
describing the so-called twisted spinors on Schwarzschild and 
Reissner-Nordstr\"om black holes and it was analysed in 
Refs. \cite{{Gon052},{Gon99}}, so we can use the results obtained 
therein for our aims. Let us adduce the necessary relations. 

The eigenvalue equation for
corresponding spinors $\Phi$ may look as follows
$${\cal D}_0\Phi=\lambda\Phi.\>\eqno({\rm A}.1)$$

As was discussed in Refs. \cite{Gon99}, the natural form of ${\cal D}_0$ 
(arising within applications) in 
local coordinates $\vartheta, \varphi$ on the unit sphere ${\Bbb S}^2$ looks 
as 
$${\cal D}_0=-i\sigma_1\left[
i\sigma_2\partial_\vartheta+i\sigma_3\frac{1}{\sin{\vartheta}}
\left(\partial_\varphi-\frac{1}{2}\sigma_2\sigma_3\cos{\vartheta}
\right)\right]=$$
$$\sigma_1\sigma_2\partial_\vartheta+\frac{1}{\sin\vartheta}
\sigma_1\sigma_3\partial_\varphi+ \frac{\cot\vartheta}{2}
\sigma_1\sigma_2         \eqno(\rm A.2)$$
with the ordinary Pauli matrices
$$\sigma_1=\pmatrix{0&1\cr 1&0\cr}\,,\sigma_2=\pmatrix{0&-i\cr i&0\cr}\,,
\sigma_3=\pmatrix{1&0\cr 0&-1\cr}\,, $$
so that $\sigma_1{\cal D}_0=-{\cal D}_0\sigma_1$.

The equation (A.1) was explored in Refs. \cite{Gon99}.
Spectrum of $D_0$ consists of the numbers
$\lambda=\pm(l+1)$              
with multiplicity $2(l+1)$ of each one, where $l=0,1,2,...$. Let us 
introduce the number $m$ such that $-l\le m\le l+1$ and the corresponding 
number $m'=m-1/2$ so $|m'|\le l+1/2$. Then the conforming eigenspinors of  
operator ${\cal D}_0$ are 
$$\Phi=\pmatrix{\Phi_1\cr\Phi_2\cr}= 
\Phi_{\mp\lambda}=\frac{C}{2}\pmatrix{P^k_{m'-1/2}\pm P^k_{m'1/2}\cr
P^k_{m'-1/2}\mp P^k_{m'1/2}\cr}e^{-im'\varphi}\> \eqno(\rm A.3) $$
with the coefficient $C=\sqrt{\frac{l+1}{2\pi}}$ and $k=l+1/2$.  
These spinors form an orthonormal basis in $L_2^2({\Bbb S}^2)$ 
and are subject 
to the normalization condition
$$\int_{{\Bbb S}^2}\Phi^{\dag}\Phi d\Omega=
\int\limits_0^\pi\,\int\limits_0^{2\pi}(|\Phi_{1}|^2+|\Phi_{2}|^2)
\sin\vartheta d\vartheta d\varphi=1\>. \eqno(\rm A.4)$$
Further, owing to the relation $\sigma_1{\cal D}_0=-{\cal D}_0\sigma_1$ we, 
obviously, have
$$ \sigma_1\Phi_{\mp\lambda}=\Phi_{\pm\lambda}\,.  \eqno(\rm A.5)$$

As to functions $P^k_{m'n'}(\cos\vartheta)\equiv P^k_{m',\,n'}(\cos\vartheta)$ 
then they can be chosen by 
miscellaneous ways, for instance, as follows (see, e. g.,
Ref. \cite{Vil91})
$$P^k_{m'n'}(\cos\vartheta)=i^{-m'-n'}
\sqrt{\frac{(k-m')!(k-n')!}{(k+m')!(k+n')!}}
\left(\frac{1+\cos{\vartheta}}{1-\cos{\vartheta}}\right)^{\frac{m'+n'}{2}}\,
\times$$
$$\times\sum\limits_{j={\rm{max}}(m',n')}^k
\frac{(k+j)!i^{2j}}{(k-j)!(j-m')!(j-n')!}
\left(\frac{1-\cos{\vartheta}}{2}\right)^j \eqno(\rm A.6)$$
with the orthogonality relation at $m',n'$ fixed
$$\int\limits_0^\pi\,{P^{*k}_{m'n'}}(\cos\vartheta)
P^{k'}_{m'n'}(\cos\vartheta)
\sin\vartheta d\vartheta={2\over2k+1}\delta_{kk'}
\>.\eqno(\rm A.7)$$
It should be noted that square of 
${\cal D}_0$ is 
$${\cal D}^2_0=-\Delta_{{\Bbb S}^2}I_2+
\sigma_2\sigma_3\frac{\cos{\vartheta}}{\sin^2{\vartheta}}\partial_\varphi
+\frac{1}{4\sin^2{\vartheta}} +\frac{1}{4}\>,
\eqno(\rm A.8)$$
while laplacian on the unit sphere is
$$\Delta_{{\Bbb S}^2}=
\frac{1}{\sin{\vartheta}}\partial_\vartheta\sin{\vartheta}\partial_\vartheta+
\frac{1}{\sin^2{\vartheta}}\partial^2_\varphi=
\partial^2_\vartheta+\cot{\vartheta}\partial_\vartheta
+\frac{1}{\sin^2{\vartheta}}\partial^2_\varphi\>,
\eqno(\rm A.9)$$
so the relation (A.8) is a particular case of the so-called 
Weitzenb{\"o}ck-Lichnerowicz formulas (see Refs. \cite{81}). 
Then from (A.1) it follows 
${\cal D}^2_0\Phi=\lambda^2\Phi$ and, when using the ansatz  
$\Phi=P(\vartheta)e^{-im'\varphi}=\pmatrix{P_1\cr P_2\cr}e^{-im'\varphi}$, 
$P_{1,2}=P_{1,2}(\vartheta)$, the equation ${\cal D}^2_0\Phi=\lambda^2\Phi$ 
turns into 
$$\left(-\partial^2_\vartheta-\cot{\vartheta}\partial_\vartheta +
\frac{m'^2+\frac{1}{4}}{\sin^2{\vartheta}}+
\frac{m'\cos{\vartheta}}{\sin^2{\vartheta}}\sigma_1\right)P=$$
$$\left(\lambda^2-\frac{1}{4}\right)P\>,
\eqno(\rm A.10)$$
wherefrom all the above results concerning spectrum of ${\cal D}_0$ can be 
derived \cite{Gon99}.

When calculating the functions $P^k_{m'n'}(\cos\vartheta)$ directly, to our 
mind, it is the most convenient to use the integral expression \cite{Vil91}

$$P^k_{m'n'}(\cos\vartheta)=\frac{1}{2\pi}
\sqrt{\frac{(k-m')!(k+m')!}{(k-n')!(k+n')!}}$$
$$\int_{0}^{2\pi}\left(e^{i\varphi/2}\cos{\frac{\vartheta}{2}}+
ie^{-i\varphi/2}\sin{\frac{\vartheta}{2}}\right)^{k-n'}\times$$
$$\left(ie^{i\varphi/2}\sin{\frac{\vartheta}{2}}+
e^{-i\varphi/2}\cos{\frac{\vartheta}{2}}\right)^{k+n'}e^{im'\varphi}d\varphi 
\eqno(\rm A.11)$$
and the symmetry relations ($z=\cos{\vartheta}$) 
$$P^k_{m'n'}(z)=P^k_{n'm'}(z), \>P^k_{m',-n'}(z)=P^k_{-m',\,n'}(z),$$ 
$$P^k_{m'n'}(z)=P^k_{-m',-n'}(z)\,,$$ 
$$P^k_{m'n'}(-z)=i^{2k-2m'-2n'}P^k_{m',-n'}(z)\>. \eqno(\rm A.12)$$
In particular
$$P^{k}_{kk}(z)=
\cos^{2k}{(\vartheta/2)},  
P^{k}_{k,-k}(z)=i^{2k}\sin^{2k}{(\vartheta/2)},$$
$$P^{k}_{k0}(z)=\frac{i^{k}\sqrt{(2k)!}}{2^k k!}\sin^{k}{\vartheta}\,,$$
$$ P^{k}_{kn'}(z)=i^{k-n'}\sqrt{\frac{(2k)!}{(k-n')!(k+n')!}}\times$$
$$\sin^{k-n'}{(\vartheta/2)}\cos^{k+n'}{(\vartheta/2)}\>. \eqno(\rm A.13)$$ 
\subsection*{Eigenspinors with $\lambda=\pm1,\,\pm2$}
If $\lambda=\pm(l+1)=\pm1$ then $l=0$ and from (A.3) it follows that 
$k=l+1/2=1/2$, $|m'|\le1/2$ and we need the functions $P^{1/2}_{m',\pm1/2}$ 
that are easily evaluated with the help of (A.11)--(A.13) so   
the eigenspinors for $\lambda=-1$ are 
$$\Phi=\frac{C}{2}\pmatrix{\cos{\frac{\vartheta}{2}}+
i\sin{\frac{\vartheta}{2}}\cr
\cos{\frac{\vartheta}{2}}-i\sin{\frac{\vartheta}{2}}\cr}e^{i\varphi/2},$$
$$\Phi=\frac{C}{2}\pmatrix{\cos{\frac{\vartheta}{2}}+
i\sin{\frac{\vartheta}{2}}\cr
-\cos{\frac{\vartheta}{2}}+i\sin{\frac{\vartheta}{2}}\cr}
e^{-i\varphi/2},\eqno(\rm A.14)$$
while for $\lambda=1$ the conforming spinors are
$$\Phi=\frac{C}{2}\pmatrix{\cos{\frac{\vartheta}{2}}-
i\sin{\frac{\vartheta}{2}}\cr
\cos{\frac{\vartheta}{2}}+i\sin{\frac{\vartheta}{2}}\cr}e^{i\varphi/2},$$ 
$$\Phi=\frac{C}{2}\pmatrix{-\cos{\frac{\vartheta}{2}}+
i\sin{\frac{\vartheta}{2}}\cr
\cos{\frac{\vartheta}{2}}+i\sin{\frac{\vartheta}{2}}\cr}e^{-i\varphi/2}
\eqno(\rm A.15) $$
with the coefficient $C=\sqrt{1/(2\pi)}$.

It is clear that (A.14)--(A.15) can be rewritten in the form 
$$\lambda=-1: \Phi=\frac{C}{2}\pmatrix{e^{i\frac{\vartheta}{2}}
\cr e^{-i\frac{\vartheta}{2}}\cr}e^{i\varphi/2},$$
or
$$\Phi=\frac{C}{2}\pmatrix{e^{i\frac{\vartheta}{2}}\cr
-e^{-i\frac{\vartheta}{2}}\cr}e^{-i\varphi/2},$$
$$\lambda=1: \Phi=\frac{C}{2}\pmatrix{e^{-i\frac{\vartheta}{2}}\cr
e^{i\frac{\vartheta}{2}}\cr}e^{i\varphi/2},$$
or
$$\Phi=\frac{C}{2}\pmatrix{-e^{-i\frac{\vartheta}{2}}\cr
e^{i\frac{\vartheta}{2}}\cr}e^{-i\varphi/2}\,, 
\eqno(\rm A.16) $$
so the relation (A.5) is easily verified at $\lambda=\pm1$. 

In studying vector mesons and excited states of heavy quarkonia eigenspinors 
with $\lambda=\pm2$ may also be useful. Then $k=l+1/2=3/2$, $|m'|\le3/2$ and we 
need the functions $P^{3/2}_{m',\pm1/2}$ 
that can be evaluated with the help of (A.11)--(A.13). Computation gives 
rise to 
$$ P^{3/2}_{3/2,-1/2}=-\frac{\sqrt{3}}{2}\sin{\vartheta}
\sin{\frac{\vartheta}{2}}= P^{3/2}_{-3/2,1/2},\>$$   
$$P^{3/2}_{3/2,1/2}=i\frac{\sqrt{3}}{2}\sin{\vartheta}
\cos{\frac{\vartheta}{2}}= P^{3/2}_{-3/2,-1/2},\>$$
$$P^{3/2}_{1/2,-1/2}= -\frac{i}{4}\left(\sin{\frac{\vartheta}{2}}-
3\sin{\frac{3}{2}\vartheta}\right)=P^{3/2}_{-1/2,1/2},\>$$
$$P^{3/2}_{1/2,1/2}= \frac{1}{4}\left(\cos{\frac{\vartheta}{2}}+
3\cos{\frac{3}{2}\vartheta}\right)=P^{3/2}_{-1/2,-1/2},\>
\eqno(\rm A.17) $$
and according to (A.3) this entails eigenspinors with $\lambda=2$ in the 
form
$$\frac{C}{2}i\frac{\sqrt{3}}{2}\sin{\vartheta}
\pmatrix{e^{-i\frac{\vartheta}{2}}\cr
e^{i\frac{\vartheta}{2}}\cr}e^{i3\varphi/2},\>
\frac{C}{8}\pmatrix{3e^{-i\frac{3\vartheta}{2}}+e^{i\frac{\vartheta}{2}}\cr
3e^{i\frac{3\vartheta}{2}}+e^{-i\frac{\vartheta}{2}}\cr}e^{i\varphi/2},\>$$
$$\frac{C}{8}\pmatrix{-3e^{-i\frac{3\vartheta}{2}}-e^{i\frac{\vartheta}{2}}\cr
3e^{i\frac{3\vartheta}{2}}+e^{-i\frac{\vartheta}{2}}\cr}e^{-i\varphi/2},\>
\frac{C}{2}i\frac{\sqrt{3}}{2}\sin{\vartheta}
\pmatrix{-e^{-i\frac{\vartheta}{2}}\cr
e^{i\frac{\vartheta}{2}}\cr}e^{-i3\varphi/2}\>
 \eqno(\rm A.18) $$
with $C=1/\sqrt{\pi}$, while eigenspinors with $\lambda=-2$ are obtained in 
accordance with relation (A.5). 

\section*{Appendix B}
We here adduce the explicit form for the radial parts of meson wave functions 
from (6). At $n_j=0$ they are given by 
$$F_{j1}=C_jP_jr^{\alpha_j}e^{-\beta_jr}\left(1-
\frac{Y_j}{Z_j}\right),$$
$$F_{j2}=iC_jQ_jr^{\alpha_j}e^{-\beta_jr}\left(1+
\frac{Y_j}{Z_j}\right),\eqno(\rm B.1)$$
while at $n_j>0$ they are given by
$$F_{j1}=C_jP_jr^{\alpha_j}e^{-\beta_jr}\times$$
$$\left[\left(1-\frac{Y_j}{Z_j}\right)L^{2\alpha_j}_{n_j}(r_j)+
\frac{P_jQ_j}{Z_j}r_jL^{2\alpha_j+1}_{n_j-1}(r_j)\right],$$
$$F_{j2}=iC_jQ_jr^{\alpha_j}e^{-\beta_jr}\times$$
$$\left[\left(1+
\frac{Y_j}{Z_j}\right)L^{2\alpha_j}_{n_j}(r_j)-
\frac{P_jQ_j}{Z_j}r_jL^{2\alpha_j+1}_{n_j-1}(r_j)\right]\eqno(\rm B.2)$$
with the Laguerre polynomials $L^\rho_{n}(r_j)$, $r_j=2\beta_jr$, 
$\beta_j=\sqrt{\mu_0^2-\omega_j^2+g^2b_j^2}$ at $j=1,2,3$ with 
$b_3=-(b_1+b_2)$, 
$P_j=gb_j+\beta_j$, $Q_j=\mu_0-\omega_j$,
$Y_j=P_jQ_j\alpha_j+(P^2_j-Q^2_j)ga_j/2$, 
$Z_j=P_jQ_j\Lambda_j+(P^2_j+Q^2_j)ga_j/2$    
with $a_3=-(a_1+a_2)$,   
$\Lambda_j=\lambda_j-gB_j$ with $B_3=-(B_1+B_2)$, 
$\alpha_j=\sqrt{\Lambda_j^2-g^2a_j^2}$, 
while $\lambda_j=\pm(l_j+1)$ are
the eigenvalues of Euclidean Dirac operator ${\cal D}_0$ 
on unit two-sphere with $l_j=0,1,2,...$ (see Appendix A) 
and quantum numbers $n_j=0,1,2,...$ are defined by the relations 
$$n_j=\frac{gb_jZ_j-\beta_jY_j}{\beta_jP_jQ_j}\,, 
\eqno(\rm B.3)$$
which entails the spectrum (5).  
Further, $C_j$ of (B.1)--(B.2) should be determined
from the normalization condition
$$\int_0^\infty(|F_{j1}|^2+|F_{j2}|^2)dr=\frac{1}{3}\>.\eqno(\rm B.4)$$
As a consequence, we shall gain that in (4) 
$\Psi_j\in L_2^{4}({\Bbb R}^3)$ at any $t\in{\Bbb R}$ and, accordingly,
$\Psi=(\Psi_1,\Psi_2,\Psi_3)$ may describe relativistic bound states 
in the field (3) with the energy spectrum (5). As is clear from (B.3) at 
$n_j=0$ we have 
$gb_j/\beta_j=Y_j/Z_j$ so the radial parts of (B.1) can be rewritten as  
$$F_{j1}=C_jP_jr^{\alpha_j}e^{-\beta_jr}\left(1-
\frac{gb_j}{\beta_j}\right),$$
$$F_{j2}=iC_jQ_jr^{\alpha_j}e^{-\beta_jr}\left(1+
\frac{gb_j}{\beta_j}\right)\>.\eqno(\rm B.5)$$
More details can be found in Refs. \cite{{Gon01},{Gon052}}. 
\section*{Appendix C}
The facts adduced here have been obained in Refs. \cite{{Gon051},{Gon052}} and 
we concisely give them only for completeness of discussion in Section 2.

To specify the question, let us note that in general the Yang-Mills equations 
on a manifold $M$ can be written as
$$d\ast F= g(\ast F\wedge A - A\wedge\ast F) \>,\eqno(\rm C.1)$$ 
where a gluonic field $A=A_\mu dx^\mu=
A^a_\mu \lambda_adx^\mu$ [$\lambda_a$ are the 
known Gell-Mann matrices, $\mu=t,r,\vartheta,\varphi$ (in the case of 
spherical coordinates), $a=1,...,8$], 
the curvature matrix (field strentgh)
$F=dA+gA\wedge A= F^a_{\mu\nu}\lambda_adx^\mu\wedge dx^\nu$ with exterior 
differential $d$ and the Cartan's (exterior) product $\wedge$, while $\ast$ 
means the Hodge star operator conforming to a metric on manifold under 
consideration, $g$ is a gauge coupling constant.

The most important case of $M$ is Minkowski spacetime and we 
are interested in the confining solutions $A$ of the SU(3)-Yang-Mills 
equations. The confining solutions were defined in Ref. \cite{Gon01} as the 
spherically symmetric solutions of the Yang-Mills 
equations (1) containing only the components of the 
SU($3$)-field which are Coulomb-like or linear in $r$. Additionally 
we impose the Lorentz condition on the sought solutions. 
The latter condition is necessary for 
quantizing the gauge fields consistently within the framework of perturbation 
theory (see, e. g. Ref. \cite{Ryd85}), so we should impose the given condition 
that can be written
in the form ${\rm div}(A)=0$, where the divergence of the Lie algebra valued
1-form $A=A_\mu dx^\mu=A^a_\mu \lambda_adx^\mu$ is defined by the relation 
(see, e. g., Refs. \cite{{Bes87}})
$${\rm div}(A)=\frac{1}{\sqrt{\delta}}\partial_\mu(\sqrt{\delta}g^{\mu\nu}
A_\nu)\>.\eqno(\rm C.2)$$
It should be emphasized that, from the physical point of view, the Lorentz 
condition reflects the fact of transversality for gluons that arise as quanta 
of SU(3)-Yang-Mills field when quantizing the latter (see, e. g., 
Ref. \cite{Ryd85}).

We shall use the Hodge star operator action on the 
basis differential 2-forms on Minkowski spacetime with local 
coordinates $t, r, \vartheta, \varphi$ in the form
$$\ast(dt\wedge dr)=-r^2\sin\vartheta d\vartheta\wedge d\varphi\>,
\ast(dt\wedge d\vartheta)=\sin\vartheta dr\wedge d\varphi\>,$$
$$\ast(dt\wedge d\varphi)=-\frac{1}{\sin\vartheta}dr\wedge d\vartheta\>,
\ast(dr\wedge d\vartheta)=\sin\vartheta dt\wedge d\varphi\>,$$
$$\ast(dr\wedge d\varphi)=-\frac{1}{\sin\vartheta}dt\wedge d\vartheta\>,
\ast(d\vartheta\wedge d\varphi)=\frac{1}{r^2\sin\vartheta}dt\wedge dr\>,
\eqno(\mathrm C.3)$$
so that on 2-forms $\ast^2=-1$. More details about the Hodge star operator can 
be found in \cite{Bes87}. 

The most general ansatz for a spherically symmetric solution is 
$A=A_t(r)dt+A_r(r)dr+A_\vartheta(r)d\vartheta+A_\varphi(r)d\varphi$. 
But then the Lorentz 
condition (C.2) for the given ansatz gives rise to 
$$\sin{\vartheta}\partial_r(r^2A_r)+
\partial_\vartheta(\sin{\vartheta}A_\vartheta)=0,$$
which yields $A_r=\frac{C}{r^2}-
\frac{\cot{\vartheta}}{r^2}\int A_\vartheta(r)dr$ with a constant matrix $C$. 
But the confining solutions should be spherically symmetric and contain only 
the components which are Coulomb-like or linear in $r$, so one should put 
$C=A_\vartheta(r)=0$. Consequently, 
the ansatz $A=A_t(r)dt+A_\varphi(r)d\varphi$ is the most general 
spherically symmetric one. 

For the latter ansatz we have $F=dA+gA\wedge A=-\partial_rA_tdt\wedge dr+
\partial_rA_\varphi dr\wedge d\varphi+g[A_t,A_\varphi]dt\wedge d\varphi$, 
where $[\cdot,\cdot]$ signifies matrix commutator.
 
Then, according to (C.3), we obtain 
$$\ast F= (r^2\sin{\vartheta})\partial_rA_td\vartheta\wedge d\varphi-
\frac{1}{\sin{\vartheta}}\partial_rA_\varphi dt\wedge d\vartheta-$$ 
$$\frac{g}{\sin{\vartheta}}[A_t,A_\varphi]dr\wedge d\vartheta\>,
\eqno(\mathrm C.4)$$ 
which entails 
$$d\ast F= 
\sin{\vartheta}\partial_r(r^2\partial_rA_t)\,dr\wedge d\vartheta\wedge d\varphi+
\frac{1}{\sin{\vartheta}}\partial_r^2A_\varphi\, dt\wedge dr\wedge d\vartheta
\>,\eqno(\mathrm C.5)$$
while 
$$\ast F\wedge A - A\wedge\ast F=
\left(r^2\sin{\vartheta}[\partial_rA_t,A_t]-
\frac{1}{\sin{\vartheta}}[\partial_rA_\varphi,A_\varphi]\right)
\,dt\wedge d\vartheta\wedge d\varphi$$
$$-\frac{g}{\sin{\vartheta}}\left([[A_t,A_\varphi],A_t]\,
dt\wedge dr\wedge d\vartheta+
[[A_t,A_\varphi],A_\varphi]\,
dr\wedge d\vartheta\wedge d\varphi\right)
\>.\eqno(\mathrm C.6)$$
Under the circumstances the Yang-Mills equations (C.1) are 
tantamount to the conditions 
$$\partial_r(r^2\partial_rA_t)=-
\frac{g^2}{\sin^2{\vartheta}}[[A_t,A_\varphi],A_\varphi],\eqno(\mathrm C.7)$$
$$\partial_r^2A_\varphi=-
{g^2}[[A_t,A_\varphi],A_t],\eqno(\mathrm C.8)$$
$$r^2\sin{\vartheta}[\partial_rA_t,A_t]-
\frac{1}{\sin{\vartheta}}[\partial_rA_\varphi,A_\varphi]=0.
\eqno(\mathrm C.9)$$
The key equation is (C.7) because the matrices $A_t, A_{\varphi}$ depend on 
merely $r$ and (C.7) can be satisfied only if the matrices 
$A_t=A_t^a\lambda_a$ and $A_\varphi=A_{\varphi}^a\lambda_a$ belong to the 
so-called Cartan subalgebra of the SU(3)-Lie algebra. Let us remind that, by definition, 
a Cartan subalgebra is a maximal abelian subalgebra in 
the corresponding Lie algebra, i. e., the commutator for any two matrices of 
the Cartan subalgebra is equal to zero (see, e.g., Ref. \cite{Bar}). For 
SU(3)-Lie algebra the conforming Cartan subalgebra is generated by the 
Gell-Mann matrices $\lambda_3, \lambda_8$ which are
$$\lambda_3=\pmatrix{1&0&0\cr 0&-1&0\cr 0&0&0\cr}\,,  
  \lambda_8={1\over\sqrt3}\pmatrix{1&0&0\cr 0&1&0\cr 
                   0&0&-2\cr}\,.\eqno(\mathrm C.10)$$
Under the situation we should have $A_t=A_t^3\lambda_3+A_t^8\lambda_8$ and 
$A_\varphi=A_{\varphi}^3\lambda_3+A_{\varphi}^8\lambda_8$, then 
$[A_t,A_\varphi]=0$ and we obtain 
$$\partial_r(r^2\partial_rA_t)=0, \partial_r^2A_\varphi=0, 
\eqno(\mathrm C.11)$$
while (C.9) is identically satisfied and (C.11) gives rise to the 
solution (3) with real constants $a_j, A_j, b_j, B_j$
parametrizing the solution which proves the uniqueness theorem of Section 2 
for the SU(3) Yang-Mills equations. 

The more explicit form of (3) is 
$$A^3_t = [(a_2-a_1)/r+A_1-A_2]/2,\>$$
$$A^8_t =[A_1+A_2-(a_1+a_2)/r]\sqrt{3}/2\>,$$
$$ A^3_\varphi = [(b_1-b_2)r+B_1-B_2]/2,$$
$$ A^8_\varphi= [(b_1+b_2)r+B_1+B_2]\sqrt{3}/2\>.\eqno(\mathrm C.12)$$

Clearly, the obtained results may be extended over all SU($N$)-groups with 
$N\ge2$ and even 
over all semisimple compact Lie groups since for them the corresponding Lie 
algebras possess just the only Cartan subalgebra. Also we can talk about the 
compact non-semisimple groups, for example, U($N$). In the latter case 
additionally to Cartan subalgebra we have centrum consisting from the matrices 
of the form $\alpha I_N$ ($I_N$ is the unit matrix $N\times N$) with arbitrary 
constant $\alpha$. 

The most relevant physical cases are of course U(1)- and SU(3)-ones 
(QED and QCD). In particular, the U(1)-case allows us to build the classical 
model of confinement (see Section 2 and Ref. \cite{GF10}). 

At last, it should also be noted that the 
nontrivial confining solutions obtained exist at any gauge coupling constant 
$g$, i. e. they are essentially {\em nonperturbative} ones.

%\section{Section title}
%\label{sec:1}
%and \cite{RefJ}
%\subsection{Subsection title}
%\label{sec:2}
%as required. Don't forget to give each section
%and subsection a unique label (see Sect.~\ref{sec:1}).
%
% For one-column wide figures use
%\begin{figure}
% Use the relevant command for your figure-insertion program
% to insert the figure file.
% For example, with the option graphics use
%\resizebox{0.75\textwidth}{!}{%
 % \includegraphics{leer.eps}}
% If not, use
%\vspace{5cm}       % Give the correct figure height in cm
%\caption{Please write your figure caption here}
%\label{fig:1}       % Give a unique label
%\end{figure}
%
% For two-column wide figures use
%\begin{figure*}
% Use the relevant command for your figure-insertion program
% to insert the figure file. See example above.
% If not, use
%\vspace*{5cm}       % Give the correct figure height in cm
%\caption{Please write your figure caption here}
%\label{fig:2}       % Give a unique label
%\end{figure*}
%
% For tables use
%\begin{table}
%\caption{Please write your table caption here}
%\label{tab:1}       % Give a unique label
% For LaTeX tables use
%\begin{tabular}{lll}
%\hline\noalign{\smallskip}
%first & second & third  \\
%\noalign{\smallskip}\hline\noalign{\smallskip}
%number & number & number \\
%number & number & number \\
%\noalign{\smallskip}\hline
%\end{tabular}
% Or use
%\vspace*{5cm}  % with the correct table height
%\end{table}
%
% BibTeX users please use
% \bibliographystyle{}
% \bibliography{}
%
% Non-BibTeX users please use

\end{document}